\documentclass[
reprint, longbibliography,
superscriptaddress,
preprintnumbers,
nobibnotes,
amsmath,amssymb,
aps, pra,
floatfix,
]{revtex4-2}

\usepackage[normalem]{ulem}
\usepackage{gensymb}
\usepackage{upgreek}
\usepackage{mathrsfs}
\usepackage{graphicx}
\usepackage{dcolumn}
\usepackage{bm}
\usepackage{floatrow}
\usepackage{float}
\usepackage[%
  colorlinks=true,
  urlcolor=blue,
  linkcolor=blue,
  citecolor=blue,
]{hyperref}
\usepackage{xcolor}
\usepackage{textcomp}
\usepackage{amsmath}
\usepackage{varwidth}
\usepackage{placeins}
\usepackage{gensymb}
\usepackage{siunitx}
\hypersetup{
    colorlinks,
    linkcolor={red!50!black},
    citecolor={blue!50!black},
    urlcolor={blue!80!black}
}
\usepackage[center]{titlesec}
  \titleformat{\section}
   {\centering\normalfont\fontsize{10}{10}\bfseries}{\thesection}{1em}{}
  \titlespacing{\section}{0pt}{12pt plus 4pt minus 2pt}{6pt plus 2pt minus 2pt}
    \titleformat{\subsection}
   {\centering\normalfont\fontsize{10}{10}\bfseries}{\thesubsection}{1em}{}
  \titlespacing{\subsection}{0pt}{12pt plus 4pt minus 2pt}{6pt plus 2pt minus 2pt}

\usepackage{tabularx}
\newcolumntype{C}{>{\centering\arraybackslash}X} 
\usepackage{multirow}

\newcommand{\unm}{Center for High Technology Materials and Department of Physics and Astronomy, University of New Mexico, Albuquerque, NM, USA}
\newcommand{\sandia}{Sandia National
Laboratories, Albuquerque, NM, USA}

\begin{document}

\title{Super-resolution diamond magnetic microscopy of superparamagnetic nanoparticles}

\date{\today}

\author{Nazanin Mosavian}
\affiliation{\unm}
\author{Forrest Hubert}
\affiliation{\unm}
\author{Janis Smits}
\affiliation{\unm}
\author{Pauli Kehayias}
\affiliation{\sandia}
\author{Yaser Silani}
\affiliation{\unm}
\author{Bryan~A.~Richards}
\affiliation{\unm}
\author{Victor~M.~Acosta}
\email{vmacosta@unm.edu}
\affiliation{\unm}

\begin{abstract}
Scanning-probe and wide-field magnetic microscopes based on Nitrogen-Vacancy (NV) centers in diamond have enabled remarkable advances in the study of biology and materials, but each method has drawbacks. Here, we implement an alternative method for nanoscale magnetic microscopy based on optical control of the charge state of NV centers in a dense layer near the diamond surface. By combining a donut-beam super-resolution technique with optically detected magnetic resonance spectroscopy, we imaged the magnetic fields produced by single $30\mbox{-}{\rm nm}$ iron-oxide nanoparticles. The magnetic microscope has a lateral spatial resolution of ${\sim}100~{\rm nm}$, and it resolves the individual magnetic dipole features from clusters of nanoparticles with interparticle spacings down to ${\sim}190~{\rm nm}$. The magnetic feature amplitudes are more than an order of magnitude larger than those obtained by confocal magnetic microscopy due to the smaller characteristic NV-nanoparticle distance within nearby sensing voxels. We analyze the magnetic point-spread function and sensitivity as a function of the microscope's spatial resolution and identify sources of background fluorescence that limit the present performance, including diamond second-order Raman emission and imperfect NV charge-state control. Our method, which uses $<10~{\rm mW}$ laser power and can be parallelized by patterned illumination, introduces a new format for nanoscale magnetic imaging.

\end{abstract}

\maketitle

\section{Introduction}
\label{sec:intro}
Magnetic microscopes based on Optically-Detected Magnetic Resonance (ODMR) spectroscopy of Nitrogen-Vacancy (NV) centers in diamond operate in a wide range of experimental conditions and offer a unique combination of spatial resolution and sensitivity. One of the first demonstrations used a single NV center to image magnetic fields in a scanning-probe format~\cite{BAL2008}. The scanning-probe method has subsequently been applied to study numerous nanomagnetic samples~\cite{MAL2012, RON2012, TET2014, RUG2015, CHA2017, ARI2018, THI2019}, with a spatial resolution down to ${\lesssim}40~{\rm nm}$~\cite{MAR2022}. Wide-field magnetic microscopy using a dense, near-surface layer of NV centers has also emerged as a powerful tool, allowing for faster imaging and higher sensitivity~\cite{GLE2015, FES2019, STE2010, PHA2011, GOU2014, SIM2016}. Both methods have enabled notable advances in the study of biology~\cite{ASL2023, ZHA2021} and materials~\cite{CAS2018, ACO2019, HO2021, MAR2022, XU2023}, but they have different drawbacks. The scanning-probe method requires a relatively complex apparatus and slow point-by-point scanning, while the widefield diamond magnetic microscope is limited by optical diffraction to a spatial resolution of ${\sim}300~{\rm nm}$~\cite{ARI2018,SCH2021}.

An intriguing alternative NV magnetic-imaging technique involves the use of far-field super-resolution microscopy methods to achieve sub-diffraction resolution~\cite{HAN2009, RIT2009, HAN2010, PEZ2010, MAU2010, WIL2011, WIL2012, ARR2013, PFE2014, YAN2014b, YAN2015, CHE2015, CHE2019, STO2021, WAN2022, JAS2017}. However, to date, most demonstrations have required either high laser power or low NV density, and were limited to the study of a small number of NV centers. One method, called Charge-State Depletion (CSD) microscopy~\cite{HAN2010, CHE2015, CHE2019, STO2021, WAN2022}, involves the use of a donut beam to drive most NV centers into their neutrally charged NV$^0$ ``dark state'', while leaving a sub-wavelength-sized core at the center in the negatively-charged NV$^-$ sensing state. This method has the benefit of using relatively low laser power and retaining high ODMR contrast for the NV$^-$ centers in the core, even when using dense near-surface layers of NV centers~\cite{CHE2019}. CSD microscopy was previously used to map changes in ODMR contrast due to the microwave field produced by proximal silver nanowires, with a feature width, ${\sim}290~{\rm nm}$ Full-Width-at-Half-Maximum (FWHM), narrower than that observed by confocal microscopy~\cite{CHE2019}.

In this paper, we demonstrate application of the CSD technique to perform super-resolution magnetic microscopy. We imaged the magnetic field produced by $30\mbox{-}{\rm nm}$ diameter super-paramagnetic iron-oxide nanoparticles (SPIONs) with a lateral spatial resolution of ${\lesssim}100~{\rm nm}$ FWHM. The improved resolution is also accompanied by a more than 10-fold increase in the magnetic feature amplitude. We identify sources of background fluorescence that limit the present performance, including diamond Raman emission and imperfect charge-state control. Our microscope uses $<10~{\rm mW}$ of laser power and can be parallelized with a multi-donut illumination scheme~\cite{CHM2013, YAN2014}, offering a path towards high-speed nanoscale magnetic microscopy.

\section{Experimental setup}
\label{sec:experimental setup}

\begin{figure*}[htbp]
\centering
\includegraphics[width=\textwidth]{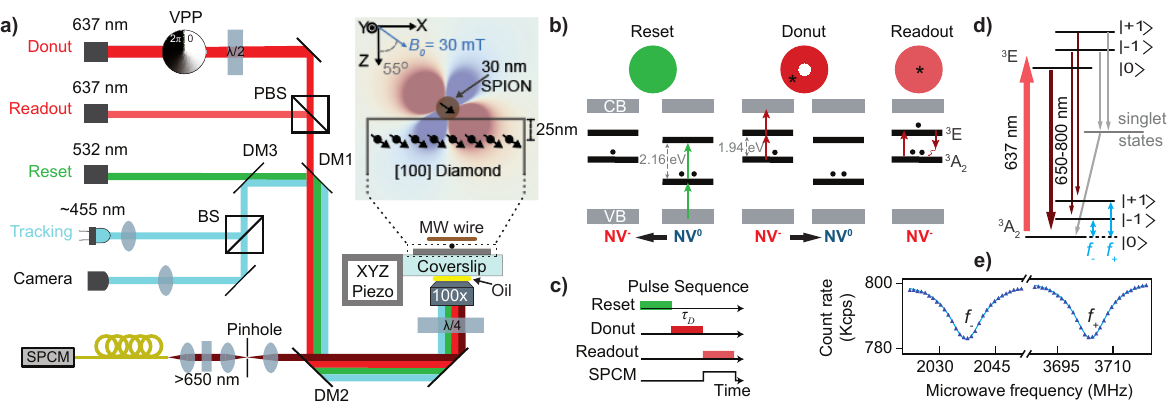}\hfill
\caption{\label{fig:experimental setup}
\textbf{Super-resolution magnetic microscopy with NV centers}.
(a) Three collimated laser beams are combined using polarizing beam splitters (PBS) and dichroic mirrors (DM). Telescopes are used to expand each beam's diameter to $\gtrsim0.5~{\rm cm}$ to match the back aperture of the objective. A vortex phase plate (VPP) is used to create the donut-shaped mode of the Donut beam. A $\lambda/4$ waveplate is used to generate the circular polarization needed to preserve the Donut beam's profile under tight focusing~\cite{GAN2003}. Emission is spatially filtered with a pinhole, spectrally filtered, and detected with an avalanche-photodiode single photon counting module (SPCM). The diamond membrane is glued to a glass coverslip, and SPION samples are dispersed on the opposite side of the diamond that hosts a dense ${\sim}25~{\rm nm}$-deep layer of NV centers. The assembly is scanned using a closed-loop piezoelectric nanopositioning stage, and a blue LED reflectance imaging system is used to minimize sample drift. Inset: depiction of the $xz$-plane magnetic field profile of a single SPION. (b) Allowed NV$^0$ and NV$^-$ optical transitions for each step of the CSD microscopy sequence. (c) Pulse sequence used for CSD microscopy. (d) Energy levels and optical and spin transitions of the NV$^-$ center. (e) Experimental confocal ODMR spectrum ($\tau_D=0$) at $B_0=30~{\rm mT}$. Lorentzian fits (blue) of the $f_{\pm}$ center frequencies are used to determine $B_{\rm nv}$.
}
\end{figure*}

The experimental apparatus is depicted in Fig.~\ref{fig:experimental setup}(a) (see also~\ref{sec:SI apparatus}). Three laser beams are combined for interrogating NV centers: two at $637~{\rm nm}$ (``Donut'' and ``Readout'') and one at $532~{\rm nm}$ (``Reset''). The laser beams are overlapped and directed to a confocal microscope with a 1.3 numerical-aperture oil-immersion objective. In the focal plane of the microscope is a dense layer of NV centers, ${\sim}25~{\rm nm}$ from the surface of a ${\sim}80\mbox{-}{\rm \upmu m}\mbox{-}$thick electronic-grade diamond membrane (\ref{sec:SI Sensor Prep}). A microwave wire is positioned on top of the diamond for driving the spin transitions of NV$^-$ centers. NV$^-$ emission is spectrally filtered (passing $>650~{\rm nm}$) and detected with a single-photon-counting avalanche photodiode. Samples are scanned using a closed-loop XYZ piezoelectric nanopositioner, and a blue LED imaging system is used to track fiducial markers milled into the diamond substrate, limiting drift to $\lesssim30~{\rm nm/day}$ ({\ref{sec:SI Tracking}}).

Figure~\ref{fig:experimental setup}(b) depicts the working principle of CSD microscopy~\cite{HAN2010, CHE2015, CHE2019, STO2021, WAN2022}. The $637{\mbox{-}}{\rm nm}$ Donut beam photo-ionizes NV centers in the high-intensity region of the beam into the NV$^0$ charge state, leaving a small sub-wavelength core at the beam's center where NV centers remain in the NV$^-$ state. The low-power, Gaussian-shaped Readout beam selectively excites NV$^-$ centers in the core, but its $637~{\rm nm}$ wavelength is too long to excite NV$^0$ centers in the periphery. The $532{\mbox{-}}{\rm nm}$ Gaussian-shaped Reset beam initializes all NV centers to an equilibrium charge-state distribution, where typically most NV centers are in the NV$^-$ state~\cite{ASL2013,CHE2017,DHO2018}. The three beams are applied in sequential $\rm \upmu s\mbox{-}scale$ pulses, Fig.~\ref{fig:experimental setup}(c), and the single-photon counting electronics only count photons during the Readout pulse (\ref{sec:SI apparatus}). 

Figure~\ref{fig:experimental setup}(d) depicts the energy levels and transitions of NV$^-$ centers that allow for magnetic field detection. A combination of optical polarization, spin-dependent fluorescence, and resonant microwave driving of the NV$^-$ spin transitions enables ODMR spectroscopy~\cite{ACO2013}. An example ODMR spectrum is shown in Fig.~\ref{fig:experimental setup}(e), revealing two peaks that are fit to Lorentzian functions. The fitted central frequencies, $f_{\pm}$, are related to the magnetic field component along the NV axis by $B_{\rm nv}\approx(f_{+}-f_{-})/(2\gamma_{\rm nv})$, where $\gamma_{\rm nv}=28.03~{\rm GHz/T}$ is the NV electron-spin gyromagnetic ratio. The magnetic field produced by the sample is then $\Delta B_{\rm nv}=B_{\rm nv}-B_0$, where $B_0=30~{\rm mT}$ is the applied bias field along the NV axis.


\section{CSD optical point spread function}
\label{sec:csd PSF}

\begin{figure*}[htbp]
\centering
\includegraphics[width=\textwidth]{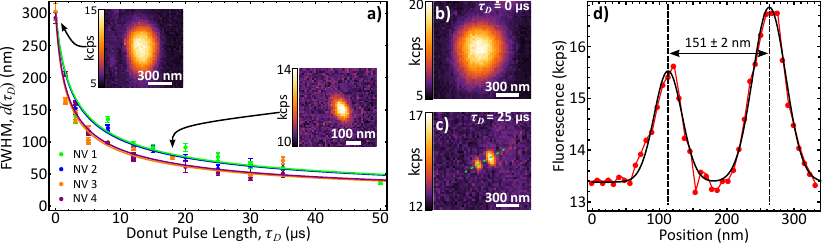}\hfill
\caption{\label{fig:CSD characterization}\textbf{CSD resolution enhancement.} (a) Lateral FWHM of CSD profiles of four individual NV centers as a function of the Donut beam pulse length. The solid colored lines are fits to Eq.~\eqref{eqn:CSD FWHM}.  The fitted values of $\tau_{\rm sat}$ for each NV center \{NV1, NV2, NV3, NV4\} are $\tau_{sat}=\{1.35\pm 0.07~{\rm \upmu s},~1.40\pm 0.14~{\rm \upmu s},~0.87\pm 0.05~{\rm \upmu s},~0.94\pm 0.07~{\rm \upmu s}\}$, respectively. Inset are two example images used in the data set. (b, c) Images of an NV cluster taken by (b) confocal microscopy ($\tau_{D}=0~{\rm \upmu s}$) and (c) CSD microscopy ($\tau_{D}=25~{\rm \upmu s}$). (d) CSD fluorescence rate along the dashed line annotated in (c). These linecut data were taken from a finer-sampled ($300~{\rm nm}\times300~{\rm nm}$) image of the same region using a $30~{\rm \upmu s}$ donut pulse length. The black solid line is a fit to two Gaussian functions, revealing an NV center separation of $151\pm2~{\rm nm}$.}
\end{figure*}

We first characterized the optical point-spread function of our CSD microscope using a diamond with a low density of NV centers located within ${\sim}1~{\rm \upmu m}$ of the surface \cite{SAN2009}. For these experiments, a $0.36\mbox{-}{\rm mW}$ Reset beam was applied for $5~{\rm \upmu s}$, a $9.8\mbox{-}{\rm mW}$ Donut beam was applied for a variable time $\tau_D$, and a $0.12\mbox{-}{\rm mW}$ Readout beam was applied for $40~{\rm \upmu s}$ (\ref{sec:SI optimize time & power}). We recorded CSD images of four isolated NV centers for different values of the donut pulse length, $\tau_D$. For each fluorescence image, we take a horizontal linecut and fit the intensity profile to a Gaussian function to extract the lateral FWHM, $d(\tau_D)$. Figure~\ref{fig:CSD characterization}(a) shows the fitted $d(\tau_D)$ for each of the 4 NV centers. In the absence of a Donut beam, the lateral FWHM is $d=296\pm7~{\rm nm}$. For a donut pulse of length $\tau_{D}=50~{\rm \upmu s}$, the lateral FWHM is $d=37\pm2~{\rm nm}$ (\ref{sec:SI best csd res}), which represents an $8$-fold resolution narrowing. At this pulse length, we also observe a ${\sim}4$-fold reduction in peak fluorescence amplitude, likely due to imperfect donut intensity contrast~\cite{SIL2019}, and a ${\sim}3$-fold increase in background fluorescence (\ref{sec:SI amp & bg}).

The data in Fig.~\ref{fig:CSD characterization}(a) are well described by the function~\cite{HAN2010,STO2021}:  %
\begin{equation}
\label{eqn:CSD FWHM}
d(\tau_D)=\frac{d_{\rm 0}}{\sqrt{1+\frac{\tau_{D}}{\tau_{\rm sat}}}},
\end{equation}
where $d_{\rm 0}=300~{\rm nm}$ is a fixed parameter representing the diffraction-limited confocal resolution and $\tau_{\rm sat}$ is a fit parameter. From the fits we extract an ensemble-averaged value $\tau_{sat}=1.04\pm 0.27~{\rm \upmu s}$ for the 4 NV centers (\ref{sec:SI amp & bg}). 

We also used our CSD microscope to resolve a pair of NV centers spaced closer than the optical diffraction limit. Figures~\ref{fig:CSD characterization}(b,c) show images of an NV cluster taken by confocal microscopy ($\tau_{D}=0~{\rm \upmu s}$) and CSD microscopy ($\tau_{D}=25~{\rm \upmu s}$), respectively. Only the CSD image is able to resolve the two NV centers. Figure~\ref{fig:CSD characterization}(d) shows a linecut from a zoomed-in CSD image. Gaussian fits reveal that the NV centers separated by $151\pm2~{\rm nm}$.

\section{Super-resolution magnetic microscopy}
\begin{figure*}[htbp]
\centering
\includegraphics[width=\textwidth]{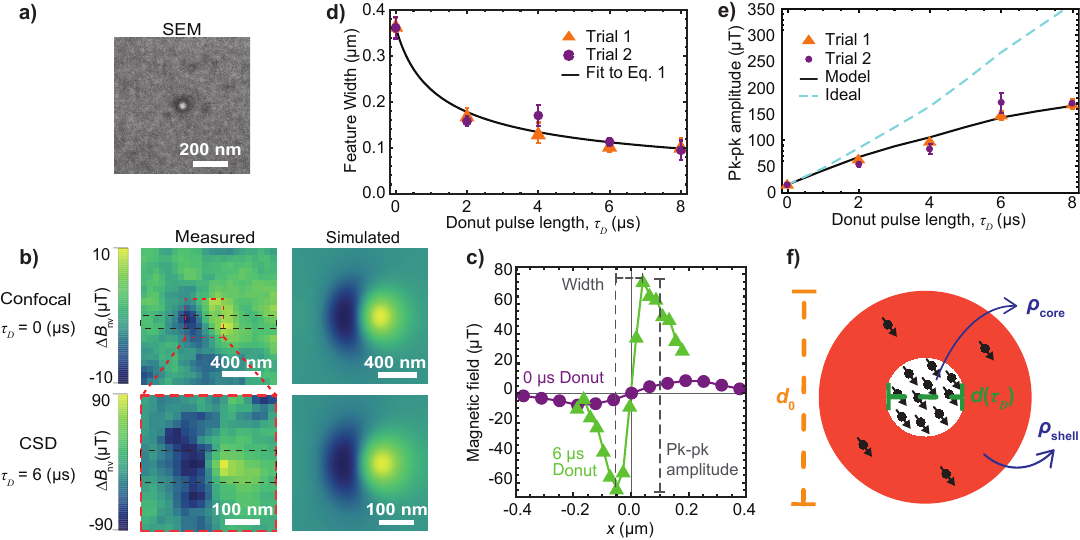}\hfill
\caption{\label{fig:mag resolution}
\textbf{Super-resolution magnetic microscopy of single SPIONs}. (a) SEM image of a single SPION on top of the diamond. (b) Top: confocal magnetic image ($\tau_D=0$) of the SPION in (a) along with a simulated profile (\ref{sec:SI Magnetic Map blur}). Bottom: CSD super-resolution magnetic image ($\tau_D=6~{\rm \upmu s}$) of the same SPION and simulated profile. (c) Horizontal linecuts through the magnetic images, as indicated in (b). The definitions of the magnetic feature width and peak-to-peak (pk-pk) amplitude are annotated with dashed lines. (d) Feature width of SPION magnetic images acquired at several values of $\tau_D$ (labeled Trial 1 and 2). For redundancy, two images were acquired for each value of $\tau_D$. The black curve is a fit to Eq.~\eqref{eqn:CSD FWHM}, with $\tau_{\rm sat}=0.63\pm0.08~{\rm \upmu s}$ and  $d_0=0.36\pm0.01~{\rm \upmu m}$. (e) Magnetic pk-pk feature amplitude as a function of $\tau_{D}$ for the same SPION magnetic images as in (d). The dashed cyan curve is the result of simulations where the SPION magnetization is constant and the NV$^-$ distribution follows the ideal behavior of Eq.~\eqref{eqn:CSD FWHM} with $\tau_{\rm sat}=0.63~{\rm \upmu s}$ (\ref{sec:SI Magnetic simulation}). The solid black curve is a model that incorporates a non-zero density of NV$^-$ centers in the high-intensity region of the Donut beam, with a shell/core density ratio $\rho_{\rm shell}/\rho_{\rm core}=0.15$. (f) Schematic of the model used for the black curve in (e). We also obtained CSD magnetic images of ferromagnetic nanoparticles, with qualitatively similar results (\ref{sec:SI micromod}).
}

\end{figure*}

We performed super-resolution magnetic microscopy of SPIONs by combining ODMR spectroscopy and CSD microscopy on a diamond with a high planar density of NV centers ($\gtrsim10^{3}~{\rm \upmu m}^{-2}$) located ${\sim}25~{\rm nm}$ from the surface (\ref{sec:SI Sensor Prep}). SPIONs form convenient test samples~\cite{GOU2014} because they effectively behave as point-like dipoles with a magnetic moment that can be controlled by moderate magnetic fields. This is due to their small size (${\sim}30~{\rm nm}$ diameter, \ref{sec:SI Sample Prep}) and superparamagnetic behavior. At $B_0=30~{\rm mT}$, the SPION magnetization is small enough that magnetic-field gradients do not meaningfully broaden the NV ODMR lines under confocal magnetic microscopy. However SPION gradients are non-negligible under super-resolution magnetic microscopy, see \ref{sec:SI contrast hole}.

A SPION suspension is drop-cast on the diamond surface at a suitable concentration such that both isolated SPIONs ($\gtrsim1~{\rm \upmu m}$ between nearest neighbors) and clusters of SPIONs ($\lesssim300~{\rm nm}$ between nearest neighbors) are present in different regions (\ref{sec:SI Sample Prep}). To correlate scanning electron microscopy (SEM) with super-resolution magnetic microscopy images, we etched fiducial markers into the diamond using focused ion beam milling (\ref{sec:SI Sensor Prep}). The fiducial markers could be resolved by SEM, fluorescence, and blue reflectance imaging. 

Figure~\ref{fig:mag resolution}(a) shows an SEM image of a single isolated SPION on the diamond surface. Figure~\ref{fig:mag resolution}(b) shows magnetic images of the same SPION with and without the Donut beam applied. Linecuts through the images are shown in Fig.~\ref{fig:mag resolution}(c). To form a magnetic image, the sample is moved to a desired position, the CSD pulse sequence in Fig.~\ref{fig:experimental setup}(c) is continuously repeated. For all magnetic imaging experiments, a $0.4{\mbox{-}}{\rm mW}$ Reset is applied for $5~{\rm \upmu s}$, a $9.8{\mbox{-}}{\rm mW}$ Donut is applied for a variable time $\tau_D$, and a $0.2{\mbox{-}}{\rm mW}$ Readout is applied for $20~{\rm \upmu s}$. Meanwhile, the microwave frequency is slowly swept (sweep time: $20~{\rm ms}$, span: $30~{\rm MHz}$)--first about the $f_{-}$ transition and then about the $f_+$ transition--to obtain ODMR spectra. The microwave sweeps are repeated and ODMR spectra are averaged together for a total pixel dwell time ($1\mbox{-}100~{\rm s}$) that is sufficient to determine $f_{\pm}$ and compute $\Delta B_{\rm nv}$ for that location. The diamond-SPION sample is then moved to the next location, and the process is repeated pixel-by-pixel until an image is formed and processed (\ref{sec:SI Magnetic Map blur}). 

With the Donut beam off ($\tau_D=0$), the SPION magnetic feature in Figs.~\ref{fig:mag resolution}(b,c) has a peak-to-peak amplitude of $14.3\pm 0.8~{\rm \upmu T}$ and a width, $366\pm 12~{\rm nm}$, that is limited by optical diffraction of the confocal microscope. With the Donut beam on ($\tau_D=6~{\rm \upmu s}$), the feature amplitude grows to $131\pm 8~{\rm \upmu T}$ and the width narrows to $111\pm 8~{\rm nm}$. 

The dramatic increase in magnetic feature amplitude observed with super-resolution magnetic microscopy is a result of the rapid decay of the SPION magnetic field with distance. We model the SPION as a point dipole, with a magnetic moment $\vec{m}$ parallel to both $\vec{B_0}$ and the relevant NV symmetry axis of our [100]-cut diamond, see Fig.~\ref{fig:experimental setup}(a). The component of the SPION magnetic field that produces shifts in the NV $f_{\pm}$ resonances is thus: 
\begin{equation}
\label{eqn:magnetic field Bnv}
\Delta B_{\rm nv} =\frac{\mu_0|\vec{m}|}{4\pi|\vec{r}|^5}(x^2-y^2+2^{3/2}xz),
\end{equation}
where $\mu_0$ is the vacuum permeability and $\vec{r}=\{x,y,z\}$ is the SPION-NV displacement vector in the lab coordinate system defined in Fig.~\ref{fig:experimental setup}(a). In simulations, we assume the NV layer is a sheet with a vertical displacement $z=40~{\rm nm}$ from the center of the SPION. The magnetic field detected by our microscope at a particular location is an average over the lateral distribution of NV$^-$ centers that contribute to the signal. Thus, even a moderate improvement in lateral spatial resolution, corresponding to a moderate reduction in characteristic values of $|\vec{r}|$, results in a large increase in the feature amplitude, see Eq.~\eqref{eqn:magnetic field Bnv}.

The observed magnetic features are reproduced by convolving Eq.~\eqref{eqn:magnetic field Bnv} with a Gaussian kernel of lateral FWHM given by Eq.~\eqref{eqn:CSD FWHM} ($\tau_{\rm sat}=0.63~{\rm \upmu s},\,d_0=0.36~{\rm \upmu m}$), 
 with only $|\vec{m}|$ as a free parameter (\ref{sec:SI Magnetic simulation}). The simulated magnetic profiles are shown to the right of the experimental data in Fig.~\ref{fig:mag resolution}(b).

\begin{figure*}[htbp]
\centering
\includegraphics[width=\textwidth]{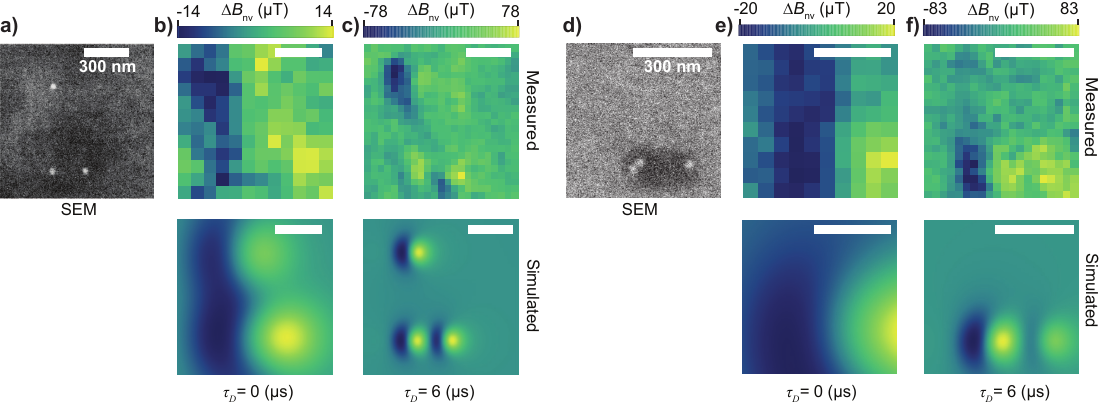}\hfill
\caption{\label{fig:multi spions}
\textbf{Super-resolution magnetic imaging of SPION clusters.} (a) SEM image of a three-SPION cluster with a smallest interparticle spacing of $226~{\rm nm}$. (b) Experimental confocal magnetic image (top) along with simulation (bottom) of the cluster in (a). (c) CSD magnetic image ($\tau_D=6~{\rm \upmu s}$) and corresponding simulation. (d) SEM of another three-SPION cluster. The dark region is due to hexane residue. (e) Experimental confocal magnetic image and corresponding simulation of the region in (d). (f) CSD magnetic image ($\tau_D=6~{\rm \upmu s}$) and corresponding simulation. Additional information about this region of the sample is included in~\ref{sec:Missed SPION}. All scale bars in (a)-(f) are $300~{\rm nm}$.}
\end{figure*}

We acquired CSD magnetic images of the same SPION for different values of $\tau_D$. Each image was processed as indicated in Fig.~\ref{fig:mag resolution}(c) to extract the feature width and amplitude. Figure~\ref{fig:mag resolution}(d) shows the fitted feature width as a function of $\tau_D$. At $\tau_D=8~{\rm \upmu s}$, the feature width is $101\pm 13~{\rm nm}$, a factor of $3.6\pm 0.5$ narrower than the confocal ($\tau_D=0$) image. The data are fit to Eq.~\eqref{eqn:CSD FWHM}, yielding $\tau_{\rm sat}=0.63\pm 0.08~{\rm \upmu s}$ and $d_0=0.36\pm0.01~{\rm \upmu m}$.

Figure~\ref{fig:mag resolution}(e) shows the feature amplitude as a function of $\tau_D$. At $\tau_D=8~{\rm \upmu s}$, the feature amplitude is $156\pm17~{\rm \upmu T}$, a factor of $11\pm 1$ increase over the confocal ($\tau_D=0$) image. The dashed cyan curve indicates the expected amplitude obtained when fixing $|\vec{m}|=0.86~{\rm A\cdot nm^2}$ at the value obtained from simulating the confocal image and setting the Gaussian blur kernel's FWHM according to the $d(\tau_D)$ fit in Fig.~\ref{fig:mag resolution}(d). 

While the amplitude observed in experiments, Fig.~\ref{fig:mag resolution}(e), increases by more than an order of magnitude from $\tau_D=0$ to $\tau_D=8~{\rm \upmu s}$, this is less than half of the expected increase. We attribute this discrepancy to imperfect control of the NV charge state, \cite{DHO2018,JAY2018,YUA2020,OLI2022,GIR2022}. As depicted in Fig.~\ref{fig:mag resolution}(f), in the high-intensity region of the donut, a small fraction of NV centers are not properly converted to NV$^0$ and exist as NV$^-$ for at least a portion of the readout pulse. The resulting CSD magnetic images are then a weighted sum of the super-resolved magnetic image and a background confocal magnetic image (\ref{sec:SI  Magnetic simulation}): 
\begin{subequations}
\label{eqn:imageweight}
\begin{align}
{\cal I}_{\rm tot}&=(1-\alpha){\cal I}_{\rm confocal}+\alpha {\cal I}_{\rm superres},\\
\alpha&\approx\frac{1}{1+\frac{\rho_{\rm shell}}{\rho_{\rm core}}\frac{\tau_D}{\tau_{\rm sat}}}.
\end{align}
\end{subequations}
Here $\rho_{\rm shell}$ is the (undesired) NV$^-$ density in the high-intensity donut ``shell'' and $\rho_{\rm core}$ is the (desired) NV$^-$ density in the donut core. Expression~\eqref{eqn:imageweight} can be qualitatively understood as follows. As $\tau_D$ increases, the area of the donut core shrinks, and thus the number of NV$^-$ centers in the core shrinks, while the number of undesired NV$^-$ centers in the shell increases. The result is that the relative weight of the super-resolution image decreases with $\tau_D$, reaching $\alpha\approx0.4$ for $\tau_D=8~{\rm \upmu s}$. By comparing the experimentally-observed amplitude with the expected values from the dashed cyan curve in Fig.~\ref{fig:mag resolution}(e), we infer a density ratio $\rho_{\rm shell}/\rho_{\rm core}\approx0.15$, see \ref{sec:SI Magnetic simulation}. The black curve in Fig.~\ref{fig:mag resolution}(e) shows the simulated amplitude with $|\vec{m}|=0.86~{\rm A\cdot nm^2}$ fixed, but allowing the relative weights of confocal and super-resolution images to vary according to Expression~\eqref{eqn:imageweight}. 

We have found that imperfect charge-state control is a particularly acute problem when using wavelengths shorter than $637~{\rm nm}$ for the Readout beam. For example, with a $610~{\rm nm}$ Readout (still longer than the ${\sim}590~{\rm nm}$ wavelength used in previous CSD microscopy studies~ \cite{CHE2015,LI2020,CHE2019,DU2020,GAR2022}), at $\tau_D=8~{\rm \upmu s}$, we realize only a ${\sim}2$-fold narrowing in feature width and a ${\sim}4$-fold increase in amplitude compared to the confocal image. 
This effect may be attributed to the shorter Readout wavelengths converting NV$^0$ to NV$^-$ through weak anti-Stokes excitation, leading to an increase in $\rho_{\rm shell}/\rho_{\rm core}$. CSD ODMR spectra obtained at different readout wavelengths further confirm this effect, see~\ref{sec:SI multi wavelength}. Our results indicate that a Readout wavelength below $637~{\rm nm}$ may be suboptimal for CSD microscopy with a high density of NV centers. Further improvements in charge-state control will benefit from continued studies of the rich interplay of diamond sensor preparation and charge-state dynamics~\cite{DHO2018,JAY2018,YUA2020,OLI2022,GIR2022}.

\section{Resolving SPION clusters}

Next, we performed super-resolution magnetic microscopy on clusters of SPIONs to demonstrate sub-diffraction spatial resolution. Figures~\ref{fig:multi spions}(a,d) show SEM images of two different clusters, each containing three SPIONs. Confocal magnetic images of each region are shown in Figs.~\ref{fig:multi spions}(b,e), along with their respective simulated profiles. CSD magnetic images ($\tau_D=6~{\rm \upmu s}$) of each region are shown in Figs.~\ref{fig:multi spions}(c,f), along with their respective simulated profiles. The simulated magnetic profiles use the same parameters as in Fig.~\ref{fig:mag resolution}(b), with the relative dipole locations determined from the SEM images, and $|\vec{m}|$ is assumed to be the same for each SPION. For both regions, the confocal magnetic images contain a single broad magnetic feature, and individual SPION features are not resolved. In the first region, Figs.~\ref{fig:multi spions}(a-c), the CSD magnetic image contains three separate SPION features that are co-localized with the SEM image, with inter-particle spacing down to $226~{\rm nm}$. In the second region, Figs.~\ref{fig:multi spions}(d-f), the CSD magnetic image contains two co-localized SPION features spaced ${\sim}190~{\rm nm}$ apart. A very close pair of SPIONs spaced $37~{\rm nm}$ apart, is still not resolved. In both regions, the CSD magnetic image feature locations are in good agreement with the SEM image, as seen in the simulations.

\section{Magnetic sensitivity}
\label{sec:mag sensitivity}

The CSD magnetic images of SPION clusters in Fig.~\ref{fig:multi spions}(c,f) took approximately one day each to acquire. Such long averaging times were required due to a relatively poor signal-to-noise ratio that degraded unexpectedly quickly with increasing $\tau_D$. To investigate further, we studied the ODMR parameters relating to magnetic sensitivity as a function of $\tau_D$. The sensitivity is fundamentally limited by photon shot noise, with a minimum detectable magnetic field given by:
\begin{equation}
\label{eqn:Bmin}
B_{\rm PSN} = \frac{4}{3 \sqrt{3} } \frac{\Gamma}{\gamma_{\rm NV} C \sqrt{I_0 \tau_{\rm avg}}}.
\end{equation}
Here ${\Gamma}$ is the ODMR FWHM, $C$ is the contrast, $I_0$ is the fluorescence count rate, and $\tau_{\rm avg}$ is the total acquisition time. Equation~\eqref{eqn:Bmin} is derived assuming the microwave frequency is tuned to the steepest slope of a Lorentzian ODMR peak. In our experiments, the microwave frequency is less-efficiently swept across the ODMR peaks, but the same scaling with $\Gamma$, $C$, and $I_0$ still applies~\cite{FES2019}.

\begin{figure}[!t]
\centering
\includegraphics[width=.95\columnwidth]{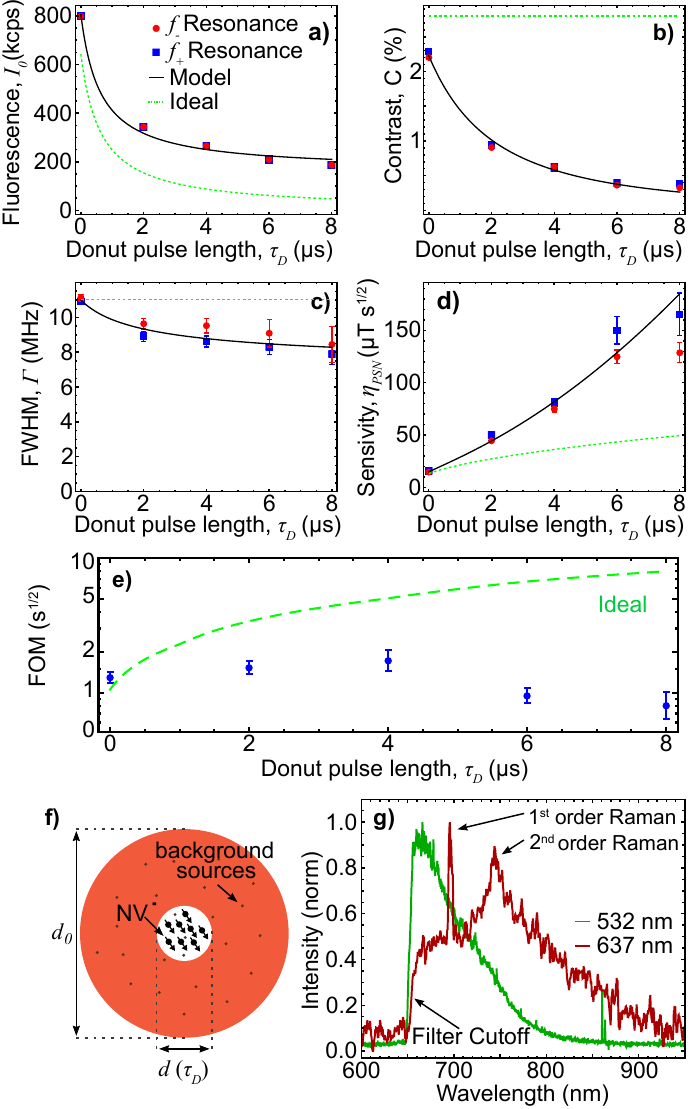}\hfill
\caption{\label{fig:sensitivity}\textbf{Magnetic sensitivity.} (a) Fluorescence rate, $I_0$, as a function of the donut pulse length, $\tau_{D}$, for both $f_{\pm}$ ODMR resonances. (b) ODMR contrast, $C$, versus $\tau_{D}$. (c) ODMR FWHM linewidth, $\Gamma$, versus $\tau_{D}$. (d) Photon-shot-noise-limited sensitivity, calculated from Eq.~\eqref{eqn:Bmin}, versus $\tau_{D}$. (e) $FOM$ versus $\tau_D$. Blue points are calculated from experiments and the dashed green curve is the simulated ideal case. In (a)-(d), solid black lines are a fit to a model that incorporates non-NV$^-$ background, depicted in (f), as well as the light-narrowing effect (\ref{sec:SI Sensitivity}). In (a)-(e), the green dashed line is an extrapolation of the model fit assuming no background or light narrowing. (g) Fluorescence spectrum obtained with either $532~{\rm nm}$ (green) or $637~{\rm nm}$ (red) excitation. The annotated 1st and 2nd order diamond Raman features appear only in the spectrum with $637~{\rm nm}$ excitation.
}
\end{figure}

The sample was positioned in a SPION-free region, and ODMR spectra were acquired about each $f_\pm$ transition for different values of $\tau_D$. Each spectrum was fit to a Lorentzian function to extract $\Gamma$, $C$, and $I_0$. Fig.~\ref{fig:sensitivity}(a) shows the peak fluorescence rate, $I_0$, as a function of $\tau_D$. A monotonic decrease in fluorescence rate is observed. This qualitative behavior is expected because as $\tau_D$ increases, the number of NV$^-$ in the donut core decreases. The observed behavior of $I_0(\tau_D)$ can be approximately described by the function:
\begin{equation}
\label{eqn:PLdecay}
I_0(\tau_D)=I_0(0)\frac{d(\tau_D)^2}{d_0^2}+bg=\frac{I_0(0)}{1+\frac{\tau_D}{\tau_{\rm sat}}}+bg,
\end{equation}
where $bg$ is a constant non-NV$^-$ background, and in the last step we used Eq.~\eqref{eqn:CSD FWHM}. Fitting to the experimental data in Fig.~\ref{fig:sensitivity}(a), and applying $\tau_{\rm sat}=0.63~{\rm \upmu s}$ from Fig.~\ref{fig:mag resolution}(d), we find $I_0(0)=642\pm19~{\rm kilocounts\,per\,second\,(kcps)}$ and $bg=162\pm9~{\rm kcps}$.

While the background fluorescence rate is relatively small compared to the NV$^-$ fluorescence rate at $\tau_D=0$, it nevertheless has a dramatic impact on the ODMR contrast for larger values of $\tau_D$. Figure~\ref{fig:sensitivity}(b) shows the ODMR contrast, $C$, as a function of $\tau_D$. At $\tau_D=0$, the ODMR contrast is $C(0)=2.2\pm 0.1\%$, a fairly typical value for our high-NV-density, ion-implanted diamonds~\cite{KEH2017,FES2019}. However, the ODMR contrast falls sharply with increasing $\tau_D$, eventually reaching $C(8~{\rm \upmu s})=0.36\pm 0.06\%$. This drop in contrast is largely due to the non-NV$^-$ background, which becomes the dominant contribution to the fluorescence rate for $\tau_D\gtrsim2~{\rm \upmu s}$, see Fig.~\ref{fig:sensitivity}(a). In order to fully describe our data, we found we also needed to include a term due to the ``light-narrowing'' effect of optical pumping by the Donut beam~\cite{DRE2011, JEN2013}, see \ref{sec:SI Sensitivity}. This choice is further justified by the decrease in $\Gamma$ as a function of $\tau_D$, Fig.~\ref{fig:sensitivity}(c), which is well described by the light-narrowing model. Additional details of the non-NV$^-$ background and ODMR parameter modeling are in \ref{sec:SI Sensitivity}.

Using Eq.~\eqref{eqn:Bmin} and the parameters from Figs.~\ref{fig:sensitivity}(a-c), we calculated the photon-shot-noise-limited sensitivity, $\eta_{\rm PSN}=B_{\rm PSN}\sqrt{\tau_{\rm avg}}$. The results are plotted in Fig.~\ref{fig:sensitivity}(d). The sensitivity degrades with increasing $\tau_D$ due to a combination of the (expected) decrease in $I_0$ and the (relatively unexpected) decrease in $C$ arising from the non-NV$^-$ background. A figure of merit for the signal-to-noise ratio in nanoparticle detection is defined as $FOM=A/\eta_{\rm PSN}$, where $A$ is the magnetic feature amplitude from Fig.~\ref{fig:mag resolution}(e). Figure~\ref{fig:sensitivity}(e) shows $FOM$ as a function of $\tau_D$ for the experimental data, along with the ideal case assuming no background fluorescence. In the present experiments, the $FOM$ hardly changes, as the increase in feature amplitude with increasing $\tau_D$ is mostly offset by the degradation in sensitivity. However, the ideal case predicts a ${\sim}8$-fold increase in $FOM$ should be possible for $\tau_D=8~{\rm \upmu s}$ compared to the confocal case.

Our observations point to the need for minimizing sources of background fluorescence in future CSD experiments, as even a low background volume intensity can become substantial when the donut core is small, as depicted in Fig.~\ref{fig:sensitivity}(f). As a first step, we acquired fluorescence spectra under intense illumination to determine the nature of the background sources. Figure~\ref{fig:sensitivity}(g) shows spectra under $532~{\rm nm}$ and $637~{\rm nm}$ illumination. For $532~{\rm nm}$ illumination, the fluorescence spectrum appears as the typical broad phonon sideband of NV centers. However, under intense, continuous-wave $637~{\rm nm}$ illumination, spectral features due to Raman emission of the diamond crystal~\cite{SOL1970} appear prominently. This suggests that diamond Raman emission contributes to the background observed in CSD ODMR spectra that limited our microscope's sensitivity (\ref{sec:SI amp & bg}). While the first-order Raman emission is sharp and could be spectrally filtered out, the second-order Raman emission is broad and covers much of the NV$^-$ fluorescence collection band. Future experiments may benefit from time-gated fluorescence filtering~\cite{FAK2008}, exploiting the large difference between the diamond Raman emission lifetime (${\lesssim}10~{\rm ps}$~\cite{Lee2010}) and NV$^-$ fluorescence lifetime (${\sim}10~{\rm ns}$~\cite{COL1983}).

\section{Discussion and conclusions}
\label{sec:discussion and conclusion}
Our results represent an early step in the application of super-resolution microscopy methods to magnetic imaging and quantum sensing. Future improvements in spatial resolution and sensitivity can be realized by eliminating the non-NV$^-$ background with time-gating~\cite{FAK2008} and reducing the undesired background NV$^-$ fluorescence through improved diamond preparation and optical charge-state control~\cite{DHO2018,JAY2018,YUA2020,OLI2022,GIR2022}. 

However these improvements alone could only, at best, render the present technique on par with single-NV scanning probe methods. In order to realize comparable sensitivity to scanning probe methods over a continuously-sampled image, ${\gtrsim}10$ NV centers per sensing voxel are required to maintain reasonable pixel-to-pixel uniformity and overcome the reductions in contrast and fluorescence collection efficiency typical of NV-ensemble measurements~\cite{TET2016}. At an optimistic NV density of $10^{12}~{\rm cm}^{-2}$~\cite{KLE2016}, this implies a minimum pixel area of ${\gtrsim}(30~{\rm nm})^2$, which is not a major improvement in spatial resolution over present scanning-probe methods~\cite{MAR2022}. 

We expect the biggest improvements to come from parallelizing the image acquisition. As a relatively-low-power, far-field super-resolution method, our technique is amenable to structured illumination methods that create arrays of thousands of donut-like modes over a wide field of view~\cite{CHM2013, YAN2014}, which could potentially provide a $\gtrsim4$ orders of magnitude speed-up in image acquisition. 

With these improvements, high-speed nanoscale magnetic imaging over a wide field of view is within reach. Such a tool could enable a broad range of new applications in materials science and biology. Examples of the latter include time-resolved studies of: formation dynamics of malarial hemozoin in response to chemical environment~\cite{FES2019}, mechanical deformations of DNA in magnetic tweezer arrays~\cite{KAZ2021}, or \textit{in vitro} magnetic screening of SPIONs and smaller magnetic nanoparticles~\cite{DEM2011,MAH2011}, just to name a few.

In summary, we demonstrated super-resolution magnetic microscopy based on charge-state depletion of diamond NV centers. We imaged the magnetic fields produced by $30\mbox{-}{\rm nm}$ diameter SPIONs with a lateral spatial resolution of ${\sim}100~{\rm nm}$ and a ${\sim}10$-fold increase in magnetic feature amplitude compared to confocal microscopy. With future improvements in Raman background suppression, NV charge-state control, and multi-donut parallelization, our technique offers a path towards high-speed nanoscale magnetic microscopy.

\begin{acknowledgments}
We gratefully acknowledge advice and support from L.~Goncalves-Webster, N.~Ristoff, M.~D.~Aiello, A.~Berzins, M.~Saleh Ziabari, A.~Laraoui, J.~T.~Damron, A.~Jarmola, A.~S.~Backer, and K.~A.~Lidke.\\
\textbf{Competing interests.} The authors declare no competing financial interests.\\
\textbf{Author contributions.} N.~M., P.~K., and V.~M.~A. conceived the idea and designed the experiments. F.~H. and N.~M. built the apparatus, acquired and analyzed data, and wrote the initial manuscript draft. B.~A.~R., Y.~S., P.~K., and J.~S. assisted in sample preparation, apparatus optimization, and data interpretation. J.~S. wrote the control software. V.~M.~A. supervised the project. All authors helped edit the manuscript. \\
\textbf{Funding.} This work was supported by the National Science Foundation (DMR-1809800, CHE-1945148, OIA-1921199), National Institute of Health (DP2GM140921), and Department of Energy (LDRD University 218242).
\end{acknowledgments}

\clearpage
\appendix
\setcounter{equation}{0}
\setcounter{section}{0}
\setcounter{table}{0}
\makeatletter
\renewcommand{\thetable}{A\arabic{table}}
\renewcommand{\theequation}{A\Roman{section}-\arabic{equation}}
\renewcommand{\thefigure}{A\arabic{figure}}
\renewcommand{\thesection}{Appendix~\Roman{section}}
\makeatother

\section{Microscope apparatus}
\label{sec:SI apparatus}

A detailed diagram of the CSD magnetic microscope is shown in Fig.~\ref{fig:experimental setup}(a) in the main text. Here we provide additional details. The Donut laser source is a Toptica iBEAM-smart-640-s. In order to generate the donut shape we pass a collimated Gaussian-shaped beam through a 629.2 nm vortex phase plate (VPP-1a, VIAVI Solutions, formerly RPC Photonics). The Readout laser source is a MRL-III-640 (CNI Laser) with a maximum power of 50 mW. A bandpass filter (FF01-637/7-25, Semrock) is used to remove any amplified spontaneous emission. Waveplates are used to ensure the Readout and Donut beams have orthogonal linear polarization. A polarizing beam splitter (PBS202, Thorlabs) is used to combine the Readout and Donut beams. The Reset laser source is a pigtailed laser diode (LP520-SF15, Thorlabs). The Reset beam is combined with Donut and Readout beams using a dichroic mirror (DMLP550R, Thorlabs). To generate ${\rm \upmu s}$-scale pulses, the Donut uses current modulation embedded in the source, and the Readout and Reset beams are controlled with Acousto-Optic Modulators (AOMs) (Brimrose TEM-85-10-532 and Gooch \& Housego 3110-120, respectively). 

The Reset, Readout, and Donut beams each have their own telescopes used to generate collimated beams with a diameter comparable to the ${\sim}6\mbox{-}{\rm mm}$ back aperture of the microscope objective. After combination, all beams pass through a quarter waveplate (WPQ10ME-633, Thorlabs) that generates circular polarization with the correct handedness needed to avoid wavefront distortions of the Donut beam from high-NA focusing~\cite{GAN2003}. The laser beams are then focused with a 100X, NA=1.3, oil immersion objective lens (1-U2B53522, Olympus). The same objective collects fluorescence, which then passes through a dichroic mirror (ZT640rdc, Chroma) and is focused by a $200\mbox{-}{\rm mm}$-focal-length tube lens (ITL200, Thorlabs) onto a pinhole. For experiments with single NV centers (Fig.~\ref{fig:CSD characterization}), we used a $75\mbox{-}{\rm \upmu m}$-diameter pinhole, and for all other experiments a $100\mbox{-}{\rm \upmu m}$-diameter pinhole (P100K, Thorlabs) was used. Light exiting the pinhole is re-collimated with a lens and passed through a $650~{\rm nm}$ longpass filter (FELH0650, Thorlabs) to isolate NV$^-$ phonon-sideband emission. A final lens focuses the light into a multi-mode fiber (M31L01, Thorlabs), which is connected to an avalanche photodiode (SPCM-AQRH-13-FC, Excelitas) used for photon counting. 

The photodetector output is connected to the counter input of a data acquisition card (NI USB-6363, National Instruments). Three-dimensional scanning of the sample is achieved by a piezoelectric nanopositioning stage (MAX311D, Thorlabs). The sample scanning is synchronized with the photon counter, via the same data acquisition card, to form images. A home-built $\rm LabVIEW$ program controls the entire sequence.

For ODMR spectroscopy, the magnetic field is generated by a permanent magnet aligned to one of the NV orientations in the diamond sample. A copper wire placed above the diamond sensor is used to deliver microwaves. A microwave signal generator (Stanford Research Systems, SG384) is used to sweep microwaves about each ODMR transitions ($f_-\approx2046~{\rm MHz}$, $f_+\approx3699~{\rm MHz}$). The microwaves are amplified (ZHL16W-43-S+, Mini-circuits) prior to the connection with the copper loop used for delivery. The data acquisition card and $\rm LabVIEW$ program synchronized the timing of the microwave sweep with the optical pulses, photon counting, and nanopositioning stage.

\section{Sensor preparation}
\label{sec:SI Sensor Prep}
The diamonds used here are electronic-grade, [100]-polished substrates that were implanted with ions. After ion implantation, they were cleaned in a tri-acid mixture (1:1:1, nitric:perchloric:sulfuric acids) at $200\degree\,{\rm C}$ for ${\sim}8~{\rm hours}$. They were subsequently annealed at $800\degree\,{\rm C}$ for $4~{\rm  hours}$ followed by $1100\degree\,{\rm C}$ for $2~{\rm hours}$~\cite{KEH2017}.

For single NV experiments (Fig.~\ref{fig:CSD characterization}), we used a diamond (called ``ME~1'' in Ref.~\cite{SIL2019}) that had a native nitrogen density $\lesssim5~{\rm ppb}$ and was implanted with Si ions at a relatively low dose, $3\times10^9~{\rm ions/cm^2}$, producing vacancies. The vertical distribution of NV centers formed after annealing is then limited by vacancy mobility, but is expected to be contained within $\lesssim1~{\rm \upmu m}$ of the surface \cite{SAN2009, ORW2012}.

\begingroup
\setlength{\tabcolsep}{8pt}
\renewcommand{\arraystretch}{1.1}
\begin{table}[!hb]
\centering
\begin{tabularx}{\textwidth}{c| C C C C} 
\multirow{2}{*}{Sensor} & \multirow{2}{*}{Size~$(\rm{mm})$} & Dose $\rm{(Ion/cm^2)}$ & \multirow{2}{*}{Ion} & Energy (keV) \\ [0.2ex] 
\hline
\rule{0pt}{3ex}
ME 1 & $2{\times}2{\times}0.5$ & $3\times 10^{9}$ & $^{28}{\rm Si}^+$ & $100$ \\ [1ex]
\hline\rule{0pt}{3ex}
CSD 2 & $2{\times}1{\times}0.1$ & $1\times 10^{13}$ &  $^{14}{\rm N}^+$ & $25$ \\ [1ex]
\hline\rule{0pt}{3ex}
CSD 3 & $2{\times}1{\times}0.1$ & $1\times 10^9$ & $^{14}{\rm N}^+$ & $25$ \\ [1ex]
\hline\rule{0pt}{3ex} 
CSD 4 & $1{\times}1{\times}0.1$ & $1\times 10^{11}$ & $^{14}{\rm N}^+$ & $25$ \\ [1ex]
\hline\rule{0pt}{3ex}
CSD 5 & $2{\times}1{\times}0.08$ & $1\times 10^{10}$ & $^{14}{\rm N}^+$ & $25$ \\ [1ex]
\hline\rule{0pt}{3ex}
CSD 6 & $2{\times}1{\times}0.08$ & $1\times 10^{13}$ & $^{14}{\rm N}^+$ & $25$ \\ 
& & $3\times 10^{12}$ & $^{14}{\rm N}^+$ & $15$ \\ [1ex]
\hline\rule{0pt}{3ex}
CSD 7 & $2{\times}1{\times}0.08$ & $1\times 10^{13}$ & $^{14}{\rm N}^+$ & $15$ \\ [1ex]
\hline
\end{tabularx}
\caption{\label{table:Diamond charachteristics}
\textbf{Parameters used for ion implantation.} ME~1 was used for single-NV studies in Fig.~\ref{fig:CSD characterization}. CSD~7 was used for magnetic microscopy data in Figs.~\ref{fig:mag resolution}, \ref{fig:multi spions}, and \ref{fig:sensitivity}. }
\end{table}
\endgroup

\begin{figure}[!ht]
\centering
\includegraphics[width=0.8\columnwidth]{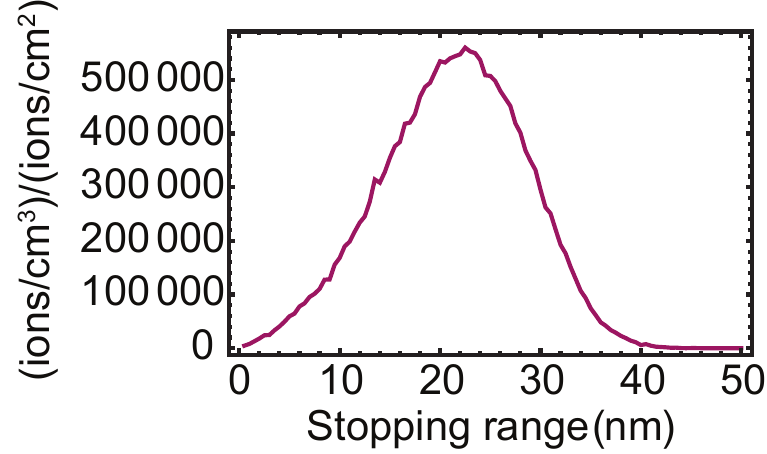}\hfill
\caption{\label{fig:SI srim}\textbf{SRIM implanted ion distribution.} Simulation for $15~{\rm keV}$ implantation of $^{14}{\rm N}^+$ in diamond at normal incidence. Simulation parameters for the diamond lattice are the same as in Ref.~\cite{KEH2017}}
\end{figure}

For magnetic microscopy, we used diamonds with a high planar density of near-surface NV centers. Electronic-grade substrates were polished to ${\sim}80~{\rm \upmu m}$ thickness and implanted with $^{14}{\rm N}^{+}$, and subsequently annealed using the recipe described above. The energy and dose of ion implantation is described in Tab.~\ref{table:Diamond charachteristics}. The diamond used for Figs.~\ref{fig:mag resolution}, \ref{fig:multi spions}, and \ref{fig:sensitivity} is called ``CSD~7''. It was implanted at a dose of $10^{13}~{\rm ions/cm^2}$ and an energy of $15~{\rm keV}$. Using Stopping Range of Ions in Matter (SRIM) simulation, Fig.~\ref{fig:SI srim}, we estimate the depth of implanted nitrogen (and thus the depth of subsequently formed NV centers) to be $\sim25~{\rm nm}$. After annealing, we estimate CSD~7 has an areal NV density of ${\sim}10^{11}~{\rm cm^{-2}}$, based on the photon count rate.

We also prepared a number of other NV-doped diamond membranes, labeled ``CSD~2-6'' in Tab.~\ref{table:Diamond charachteristics}. We used CSD~6 (two implantation energies of $15~{\rm and~25~keV}$) to perform CSD magnetic microscopy of micromod iron-oxide nanoparticles, described in \ref{sec:SI micromod}. We observed qualitatively similar behavior to the SPION experiments with CSD~7 discussed in the main text. There is a drop in ODMR contrast with increasing $\tau_D$, and there is an increase in magnetic feature amplitude and decrease in feature width of similar magnitude to that in Fig.~\ref{fig:mag resolution}. The fluorescence countrate and ODMR linewidth were also similar, at least to within a factor of ${\sim}1.3$. We used CSD~2 for ODMR spectroscopy measurements and initial SPION imaging. This diamond was slightly brighter than CSD~7, but not by more than a factor of ${\sim}1.3$. The ODMR contrast and linewidth were similar to CSD~7. Ultimately we selected CSD~7 for the results reported in the main text because the implantation depth is slightly more shallow (resulting in larger magnetic feature amplitude), while the ODMR properties did not appear to be much different.

\begin{figure}[!ht]
\centering
\includegraphics[width=0.8\columnwidth]{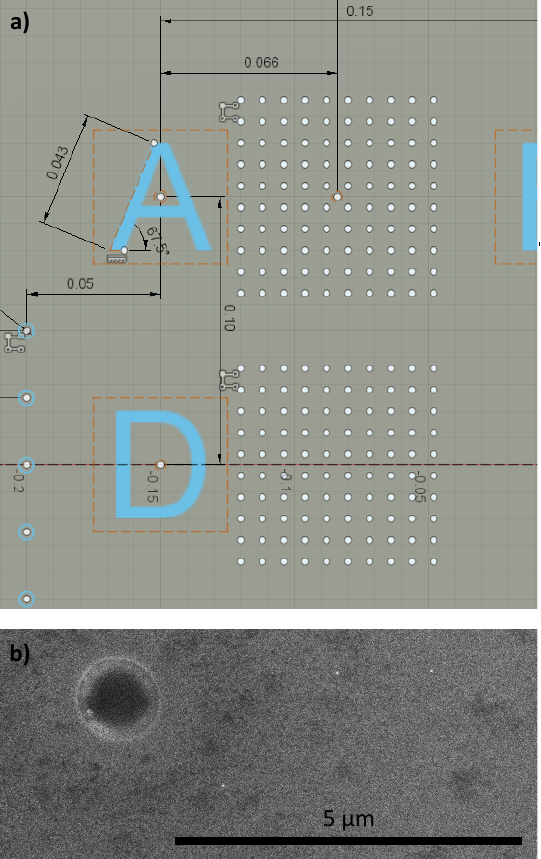}\hfill
\caption{\label{fig:SI sample prep}\textbf{Fiducial markers.} (a) CAD drawing used for FIB patterning (dimensions in mm). (b) Typical SEM image after drop-casting SPIONs onto the diamond surface. White dots are SPIONs on the diamond surface. The dark areas are due to residue from the hexane solvent (they are not present before drop-casting). There is also some additional texture near the FIB-milled hole that is due to damaged from the FIB process. We studied SPIONs in regions far enough from the markers to avoid these damaged regions.}
\end{figure}

After sensor creation, we created fiducial markers on the diamond surface that were visible under SEM, confocal fluorescence microscopy, and blue-reflectance imaging. We made these markers for two purposes. First, we needed a way to correlate magnetic microscopy images to SEM images to co-localize nanoparticles. Second, some of the fiducial markers were used for the blue-reflectance tracking and feedback, described in \ref{sec:SI Tracking}.

The markers were created using focused ion beam milling of the diamond surface through the NV-doped layer. Figure~\ref{fig:SI sample prep}(a) shows a CAD drawing of the pattern milled into the diamond. The markings have three distinct types of marks. There are large ${\sim}40\mbox{-}{\rm \upmu m}$-sized letters that are used to coarsely navigate across the surface. Next to each letter are $10\times10$ arrays of $1\mbox{-}{\rm \upmu m}$-diameter holes  (pitch of ${\sim}8~{\rm \upmu m}$) that are used for finer co-localization of nanoparticles and for blue-reflectance tracking.  Finally, near the letter ``D'', we milled donut shapes in the diamond, leaving behind narrow pillars in the middle. The pillar diameters ($50{\mbox{-}}200~{\rm nm}$) are smaller than the diffraction limit, so the ensemble of NV centers in the top of each pillar acts as a point-like source. We use these pillars to characterize the Donut, Reset, and Readout beam spatial modes and overlap.

The un-doped surface of the diamond membranes are glued to a No. 0 coverslip (CG00C2, $85\mbox{-}115~{\rm \upmu m}$ thick) using UV-curing optical adhesive (Norland 88). This allows us to focus on the NV-doped surface without exceeding the working distance of the oil-immersion objective. The refractive indices of the UV-curing adhesive, the coverglass, and the immersion oil were all approximately the same ($1.5\mbox{-}1.6$), which minimizes scattering distortions and losses.

\section{Tracking}
\label{sec:SI Tracking}
To improve the resiliency of the setup to positional drift, we implemented an active feedback mechanism. A blue reflectance channel, Fig.\ref{fig:experimental setup}(a), was added with the goal of localizing an array of fiducial markers etched into the diamond surface (\ref{fig:SI sample prep}). A low power ($\lesssim100~{\rm \upmu W}$) of light from a blue light-emitting diode (LED) was used to illuminate a region of the diamond adjacent to the magnetic microscope field of view. The reflected blue light was imaged onto a CMOS camera. While the spatial resolution of the blue imaging path is $\gtrsim250~{\rm nm}$, it was possible to localize the position of the fiducial markers to within a few nanometers. 

The localization was done in the following steps:
\begin{enumerate}
    \item The image was subdivided into four sub-regions, each containing a single $1\mbox{-}{\rm \upmu m}$-diameter-hole fiducial marker
    \item The subdivided images were passed through a canny edge detection routine implemented in $\rm LabVIEW$. 
    \item The ``center-of-mass'' (first moment) was calculated for each sub-region.
    \item The position of the diamond was computed based on the first moment of the center-of-masses of each of the individual sub-regions.  
\end{enumerate}

For CSD magnetic microscopy experiments, the $xy$ position of the first pixel in the image was set by adjusting the nanopositioning stage until the localization procedure described above reached the desired initial position on the diamond. The CSD pulse sequence was repeated and ODMR spectra were obtained and averaged for the desired pixel dwell time. Meanwhile, the localization procedure was repeated continuously in parallel. The final localization result at the end of the pixel dwell time was used to compute $xy$ step increments for the nanopositioning stage such that the next pixel would correspond to the desired next location on the diamond. For long pixel dwell times (for example $100~{\rm s}$, the localization results could drift by as much as $10~{\rm nm}$ over the course of a single pixel dwell time). By computing the step increments based on the final localization, we were able to account for this drift and steer the stage to the correct next position. 

After the magnetic imaging routine reached the end of a row, we re-positioned the $z$ axis of the stage. This was done by sweeping the $z$ position of the stage through a span of ${\sim}1~{\rm \upmu m}$ and selecting the $z$ position that gave the highest Readout fluorescence countrate. During long experiments, the drift in $z$ from row to row was relatively small--on the order of $100~{\rm nm}$--but the variations could be more severe if the lab temperature was varying.

We repeated experiments on the same SPION using the same scan parameters but on different days. We found that the SPION magnetic features always appeared in the same location to within ${\sim}\pm30~{\rm nm}$, which serves as a measure on the day-to-day accuracy of the image stabilization routine. In the absence of the localization feedback, the drift errors would accumulate, producing cumulative drifts at the level of $\gtrsim100~{\rm nm/day}$. This lead to blurring and distortions for long image acquisition times, which the image stabilization routine eliminated.

\section{Optical power and pulse lengths}
\label{sec:SI optimize time & power}
 The Reset, Donut, and Readout beam powers and pulse lengths used in the CSD pulse sequence, Fig.~\ref{fig:experimental setup}(c), were chosen based on the following considerations. First, the Readout pulse length is ideally a substantial fraction ($\gtrsim50\%$) of the overall sequence length, since only photons collected during the Readout pulse are counted. Second, to maximize sensitivity we aim to collect as many photons per second as possible without sacrificing ODMR contrast or damaging samples. Since the Readout can photo-ionize NV centers from NV$^-$ to NV$^0$, there exists an optimal pulse length, for a given Readout power, whereby at the end of the pulse $\gtrsim1/e$ of NV centers remain as NV$^-$. Third, in order to realize the best possible spatial resolution, we select a Donut power and maximum Donut pulse length that provides a high value of $\tau_D/\tau_{\rm sat}$, see Eq.~\eqref{eqn:CSD FWHM}.

 The iron-oxide nanoparticles studied here are resistant to photodamage, so we were able to choose a relatively high Donut power, $9.8~{\rm mW}$ measured after the microscope objective. The Donut pulse length needed to reach a spatial resolution $\lesssim50~{\rm nm}$ was $\tau_D\approx40~{\rm \upmu s}$ for the experiments with single NV centers (Fig.~\ref{fig:CSD characterization}). We thus set the Readout pulse length to be $40~{\rm \upmu s}$ for those experiments and tuned the Readout power such that at the end of the Readout pulse the countrate had fallen to ${\sim}1/e$ of its initial value. This resulted in a Readout power of $0.12~{\rm mW}$ for single-NV experiments. For CSD magnetic microscopy, the Donut pulse length needed to reach a spatial resolution $\lesssim100~{\rm nm}$ was $\tau_D\approx10~{\rm \upmu s}$. We thus set the Readout pulse length to be $20~{\rm \upmu s}$ for those experiments and found an optimal Readout power of $0.2~{\rm mW}$. In either case, we found that using a higher Readout power (and re-optimzing the Readout pulse length) did not result in a higher time-averaged countrate, presumably due to the nonlinear nature of the NV$^-$ to NV$^0$ photoionization, see Fig.~\ref{fig:experimental setup}(b).
 
 We found that a Reset power of $0.4~{\rm mW}$ applied for a time longer than ${\sim}2~{\rm \upmu s}$ was sufficient to revive the NV$^-$ countrate to its peak value, so we conservatively set the Reset pulse length to $5~{\rm \upmu s}$.

Future experiments with more fragile samples may select lower powers and longer pulse lengths to minimize photodamage. This will be accompanied by a modest reduction in magnetic sensitivity, but the exact scaling depends on details of the NV optical charge-state dynamics dynamics that are in turn dependent on sample preparation~\cite{DHO2018,JAY2018,YUA2020,OLI2022,GIR2022}.

\section{Best CSD resolution}
\label{sec:SI best csd res}

\begin{figure}[!ht]
\centering
\includegraphics[width=0.8\columnwidth]{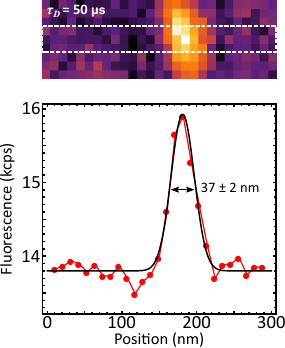}\hfill
\caption{\label{fig:best csd res}\textbf{Fluorescence and linecut of NV center imaged with 50 $\mathrm{\mathbf{\upmu s}}$ donut.} (top) The fluorescent image of an NV center taken at donut pulse length $\tau_{D}=50~\mathrm{\mu s}$. The white dashed box represented the averaged rows to make the plot below. (bottom) Linecut created by averaging the three rows in the white dashed box in the fluorescence image above. The black line is a Gaussian fit to the data, and the FWHM extracted from this fit is $37 \pm 2~{\rm nm}$.}
\end{figure}

In Fig.~\ref{fig:CSD characterization}, we reported experiments on 4 single NV centers in sample ME~1 that characterized the CSD optical point spread function. Figure \ref{fig:best csd res} shows the CSD fluorescence image and linecut used to report the best resolution we have observed with our CSD microscope. A FWHM lateral resolution of $37\pm 2~{\rm nm}$ is realized with $\tau_D=50~{\rm \upmu s}$.

\section{CSD amplitude \& background}
\label{sec:SI amp & bg}
\begin{figure}[!ht]
\centering
\includegraphics[width=0.8\columnwidth]{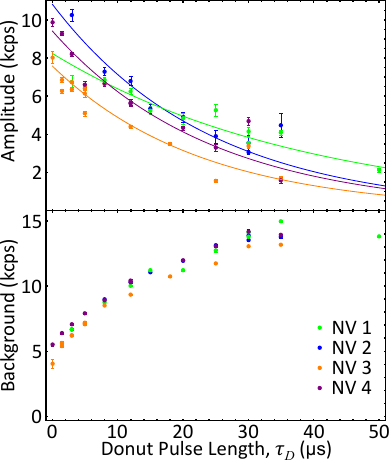}\hfill
\caption{\label{fig:NV amp & BG}\textbf{Single-NV amplitude and background.} (top) Fluorescence amplitude versus $\tau_{D}$ for four NV centers. The solid lines are fits to a decaying exponential function. The $1/e$ decay times are shown in Table~\ref{table:NV sat & amp}. (bottom) Fluorescence background versus $\tau_{D}$ for the same four NV centers.}
\end{figure}

In Fig.~\ref{fig:CSD characterization} of the main text, we reported the lateral FWHM of the CSD optical point-spread function (PSF) for four single NV centers as a function of $\tau_D$. The same images and fitting routine were also used to extract the fluorescence amplitude and background, Figure~\ref{fig:NV amp & BG}. While there is some variation among the four NV centers, we always observe a decay in the fluorescence amplitude as a function of $\tau_D$. We attribute this to imperfect intensity contrast of the Donut beam spatial profile~\cite{SIL2019}. As $\tau_D$ increases, a non-zero intensity in the Donut core photo-ionizes NV$^-$ centers to NV$^0$, reducing the countrate during the subsequent Readout. The fluorescence amplitude decay is well described by an exponential decay function, with a $1/e$ decay time, $\tau_p$, that is in the $25{\mbox{-}}40~{\rm \upmu s}$ range, see Table~\ref{table:NV sat & amp}. This effect represents a technical challenge for future optimized CSD microscopy experiments, but it did not have a substantial impact in the present CSD magnetic microscopy experiments, where $\tau_D\leq8~{\rm \upmu s}\ll\tau_p$.

\begingroup
\setlength{\tabcolsep}{8pt}
\renewcommand{\arraystretch}{1.1}
\begin{table}[hb]
\centering
\begin{tabular}{c| c c c} 
 & $\tau_{\rm sat}~({\rm\upmu s})$ & $\tau_{p}~({\rm\upmu s})$ & $\tau_{p}/\tau_{\rm sat}$ \\ [0.2ex] 
\hline
 NV 1 & $1.35\pm 0.07$ & $24\pm 2$ & $18\pm 2$ \\ 
 NV 2 & $1.40\pm 0.14$ & $39\pm 4$ & $28\pm 4$ \\
 NV 3 & $0.87\pm 0.05$ & $23\pm 3$ & $26\pm 4$ \\
 NV 4 & $0.94\pm 0.07$ & $24\pm 3$ & $26\pm 3$ \\
 Mean & $1.04\pm 0.27$ & $26\pm 8$ & $25\pm 4$ \\ [0ex]
 \hline
\end{tabular}
\caption{\label{table:NV sat & amp}\textbf{CSD optical PSF parameters.} $\tau_{\rm sat}$ is the fitted FWHM saturation time for each NV center shown in Figure~\ref{fig:CSD characterization}(a). $\tau_{p}$ is the fitted amplitude decay time for each NV center from Figure~\ref{fig:NV amp & BG}(top). $\tau_{p} / \tau_{\rm sat}$ is the ratio of the two saturation times shown in the table. For each column, the ``Mean'' is the variance-weighted average and its uncertainty is the un-weighted standard deviation of the four point estimates.}
\end{table}
\endgroup

We also observe that the fluorescence background (the countrate far away from an NV center) increases as a function of $\tau_D$, Fig.~\ref{fig:NV amp & BG} (bottom). This was unexpected, especially since in CSD magnetic microscopy experiments with high-NV-density samples, the background can be modeled as independent of $\tau_D$. Regardless, even at $\tau_D=0$, we observe a non-zero background of $\sim5~{\rm kcps}$ in Fig.~\ref{fig:NV amp & BG} (bottom). This background is likely due to second-order Raman emission (see main text), but it is more than an order of magnitude lower than the background observed indirectly in CSD magnetic microscopy experiments, see \ref{sec:SI Sensitivity}. The difference is partially attributed to the ${\sim}2$-fold smaller pinhole area and the ${\sim}2$-fold lower Readout power used in the single-NV experiments. The remaining discrepancy may be due to variation in the properties of the diamond sensor, but this remains a topic for future work.

\section{SPION sample preparation}
\label{sec:SI Sample Prep}
The SPION samples used here are $30\mbox{-}{\rm nm\mbox{-}diameter}$ particles coated with oleic acid and suspended in chloroform, purchased from Ocean NanoTech (SOR30). The stock suspension has a concentration of $25~{\rm mg/mL}$. We dilute this suspension by combining $0.2~{\rm \upmu L}$ of the stock suspension with $800~{\rm \upmu L}$ of hexane. We sonicate the diluted suspension and then drop-cast $100~{\rm \upmu L}$ of it on top of a clean diamond, using a pipette. The hexane evaporates in air almost immediately, so the result is a semi-random distribution of SPIONs on the diamond surface. We then take SEM images of regions containing single, isolated SPIONs and use the fiducial markers to signal their location for further study. Figure~\ref{fig:SI sample prep}(b) shows a typical SEM image of the diamond surface after drop-casting. Three isolated SPIONs appear as bright white dots in the image.

\section{Magnetic image processing}
\label{sec:SI Magnetic Map blur}
\begingroup
\setlength{\tabcolsep}{3pt}
\renewcommand{\arraystretch}{1.0}
\begin{table}[!hb]
\centering
\begin{tabularx}{\textwidth}{c| C C C C} 
\multirow{2}{*}{Fig.} & Pixel size (nm) & Kernel radius (pixels) & Kernel FWHM (pixels) \\ [0.2ex] 
\hline
 3b top& 66  & 22 & 2 \\ 
 3b btm& 24  & 20 & 2 \\
 4b  & 96 & 2  & 1.4    \\
 4c  & 46  & 2  & 1.4    \\
 4e  & 63  & 2  & 1.4    \\
 4f  & 31  & 2  & 1.4    \\ [0.2ex] 
 \hline
\end{tabularx}
\caption{\label{table:}
\textbf{CSD magnetic image processing.} Pixel dimensions of each experimental CSD magnetic image in the main text and the parameters of the Gaussian blur kernel applied to them.}
\end{table}
\endgroup

As described in the main text, we extract $B_{\rm nv}$ for each pixel in a CSD image by fitting the ODMR spectra to obtain $f_{\pm}$. We then convolve the magnetic images with a Gaussian kernel to provide a modest improvement in signal-to-noise ratio. This is possible to do without distorting image features because we oversample the images; in other words, we choose pixel sizes that are smaller than the lateral FWHM of the optical PSF. The FWHM of the Gaussian kernel used for convolution is chosen to be at least a factor of two narrower than the magnetic feature width to avoid distortion. This smoothing convolution is applied prior to taking the image linecuts that are used to extract feature width and pk-pk magnetic amplitude. The values used for the Gaussian kernel are shown in Table~\ref{table:}.

\section{Simulating magnetic profiles}
\label{sec:SI Magnetic simulation} 
Using Eq. \ref{eqn:magnetic field Bnv}, we calculated magnetic profiles of a SPION at distance $z=40~{\rm nm}$ from the NV center. Here $40~{\rm nm}$ is estimated from the sum of the typical depth of NV centers ($\sim25~{\rm nm}$) and the typical SPION radius ($\sim15~{\rm nm}$). As discussed in the main text, we found that if we held the SPION magnetization constant at the value that matched the confocal $\tau_D=0$ image, $|\vec{m}| = 0.86~{\rm A\cdot nm^2}$, the simulations for $\tau_D>0$ gave a magnetic feature amplitude larger than that observed in experiments. For the $\tau_D=6~{\rm \upmu s}$ simulation in Fig.~\ref{fig:mag resolution}(b), we allowed $|\vec{m}|$ to vary to match the observed amplitude, but there is no physical explanation for why $|\vec{m}|$ should depend on $\tau_D$.

As discussed in the main text, our hypothesis is that the discrepancy is due to unwanted NV$^-$ in the Donut shell adding background signal. Just underneath the SPION, NV$^-$ centers in the shell experience much smaller values of $B_{\rm nv}$ than NV$^-$ centers in the Core, leading to an overall smaller average field. The resulting CSD magnetic images are then a weighted sum of the super-resolved magnetic image and a background confocal magnetic image, see Eq.~\eqref{eqn:imageweight}. Under this hypothesis, the super-resolved image weighting factor, $\alpha$, should decrease monotonically with increasing $\tau_D$. Specifically:
\begin{equation}
    \label{eqn:derivealpha}
\begin{split}
\alpha&\approx\frac{N_{\rm core}}{N_{\rm core}+N_{\rm shell}}\\
&=\frac{\rho_{\rm core}A_{\rm core}}{\rho_{\rm core}A_{\rm core}+\rho_{\rm shell}A_{\rm shell}}\\
&=\frac{\rho_{\rm core}\frac{\pi}{4}\frac{d_0^2}{1+\tau_D/\tau_{\rm sat}}}{\rho_{\rm core}\frac{\pi}{4}\frac{d_0^2}{1+\tau_D/\tau_{\rm sat}}+\rho_{\rm shell}(\frac{\pi}{4}d_0^2-\frac{\pi}{4}\frac{d_0^2}{1+\tau_D/\tau_{\rm sat}})}\\
&=\frac{1}{1+\frac{\rho_{\rm shell}}{\rho_{\rm core}}\frac{\tau_D}{\tau_{\rm sat}}}.
\end{split}
\end{equation}
Here, $N_{\rm core,shell}$ are the number of NV centers in the core and shell, respectively, $A_{\rm core,shell}$ is the area of the core and shell, respectively, and we used Eq.~\eqref{eqn:CSD FWHM}. 

\begin{figure}[ht]
\centering
\includegraphics[width=0.8\columnwidth]{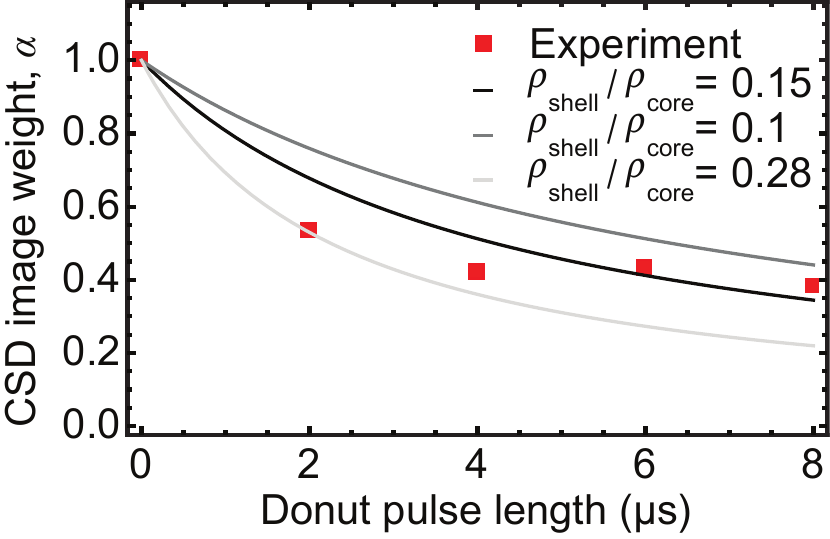}\hfill
\caption{\label{fig:SI alpha}
\textbf{CSD image weight.} Extracted values of $\alpha$ (red squares) are plotted as a function of $\tau_D$. The solid lines use weights from Eq.~\eqref{eqn:derivealpha} for different values of the NV$^-$ density ratio $\rho_{\rm shell}/\rho_{\rm core}$. The value $\rho_{\rm shell}/\rho_{\rm core}=0.15$ is the one used to generate the black curve in Fig.~\ref{fig:mag resolution}(e).}
\end{figure}

\begin{figure*}[ht!]
    \centering
    \includegraphics[width=0.7\textwidth]{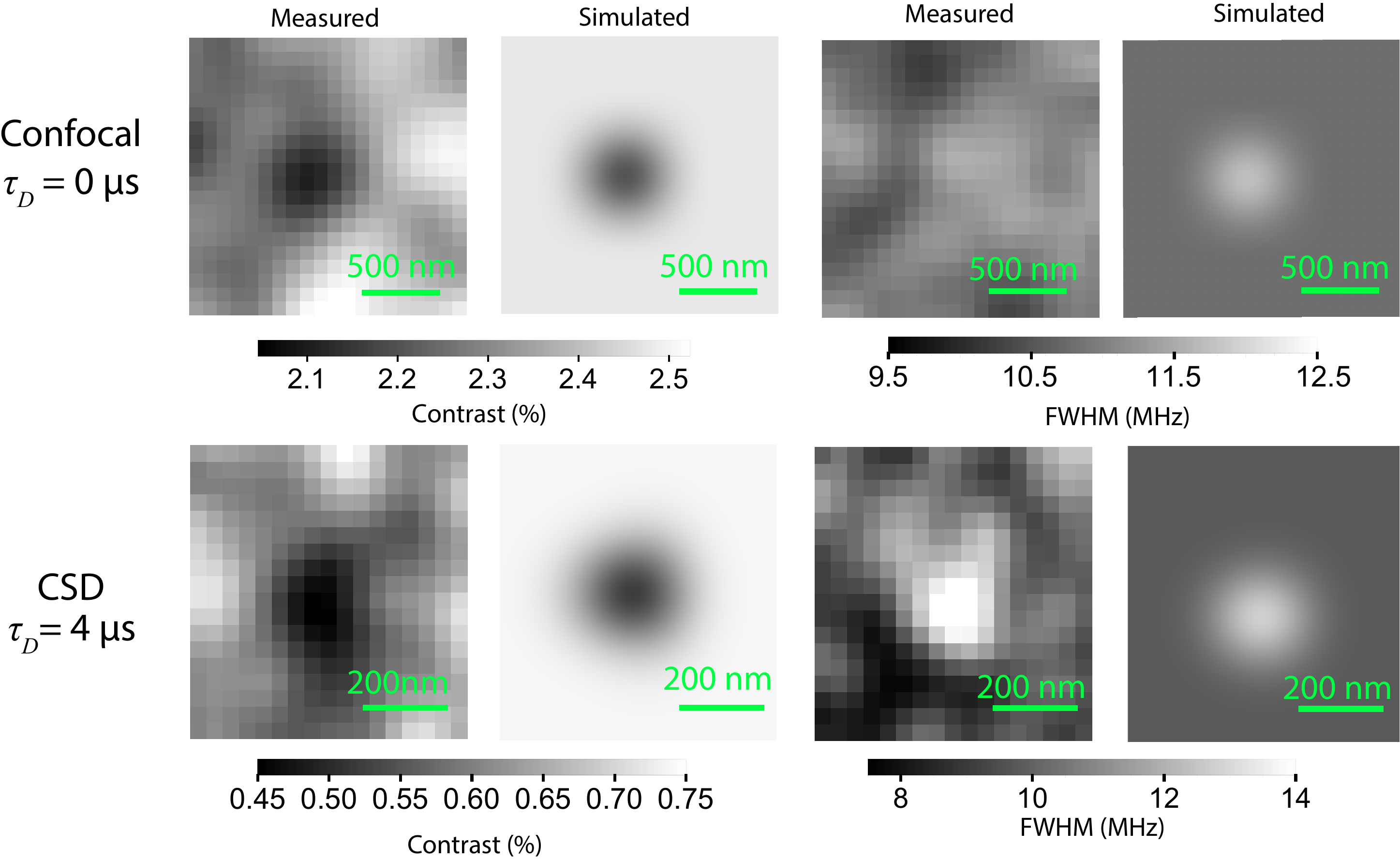}
    \caption{\textbf{SPION magnetic gradients.} (Top) Experimental and simulated ODMR contrast and FWHM for confocal imaging (PSF FWHM $\sim360$ nm). (Bottom) Contrast and FWHM for CSD microscopy with $\tau_D=4~{\rm \upmu s}$ (PSF FWHM $\sim 135$ nm).}
    \label{fig:sup10}
\end{figure*}

To check this hypothesis, for each value of $\tau_D$, we adjusted the relative weights of the confocal and super-resolved simulated profiles until the magnetic feature amplitude  matched the experimentally-observed value. The resulting ``experimental'' image weights are plotted versus $\tau_D$ in Fig.~\ref{fig:SI alpha}.
Using the independently-measured $\tau_{\rm sat}=0.63~{\rm \upmu s}$, the only dependent variable in the last line of Eq.~\eqref{eqn:derivealpha} is the NV$^-$ density ratio, $\rho_{\rm shell}/\rho_{\rm core}$. We calculated $\alpha(\tau_D)$ for different values of $\rho_{\rm shell}/\rho_{\rm core}$, shown as solid lines in Fig.~\eqref{fig:SI alpha}. The experimental weights are consistent with a range of values of $\rho_{\rm shell}/\rho_{\rm core}=0.1\mbox{-}0.28$. It is possible that the density ratio might change as a function of $\tau_D$, but we find that holding $\rho_{\rm shell}/\rho_{\rm core}=0.15$ fixed still gives fairly good agreement with the experimental feature amplitudes, as seen in Fig.~\ref{fig:CSD characterization}(e). This density ratio can be thought of as an upper bound on the error of the charge-state control. The probability for an NV center in the Donut shell to be in the NV$^-$ charge state during Readout is on the order of $0.15$, instead of the desired probability of 0.

\section{SPION magnetic field gradient}
\label{sec:SI contrast hole}
NV centers that are located close to the SPION (distance $\ll300~{\rm nm}$) experience SPION magnetic fields that are large enough to shift their $f_{\pm}$ ODMR frequencies by a substantial fraction of the ODMR natural linewidth (typically $\sim10~{\rm MHz}$ for the high-NV-density, shallow-implanted diamond sensors used here). At such a close proximity, the SPION magnetic field also varies rapidly with distance. The combined effect can lead to a broadening of the ensemble ODMR spectrum and an associated loss in contrast near the SPION. In confocal or widefield magnetic microscopy this is usually not an issue since most NV centers within a diffraction-limited PSF are located a significant distance away from the SPION and experience only a modest field. However in super-resolution magnetic microscopy, the PSF is much smaller and the fluorescence contribution of NV centers in close proximity to the SPION cannot be neglected.

Figure~\ref{fig:sup10} shows experimental $xy$ maps of the ODMR FWHM and contrast for the confocal ($\tau_D=0$) and CSD ($\tau_D=4~{\rm \upmu s}$ cases. For the confocal case, the experimental data shows a small drop in contrast near the SPION, from ${\sim}2.4\%$ away from the particle to ${\sim}2.1\%$ directly underneath. Any corresponding feature in FWHM is obscured by pixel-to-pixel variation. However, for $\tau_D=4~{\rm \upmu s}$, where the lateral PSF FWHM is $\sim135~{\rm nm}$, a large dip in contrast and increase in in FWHM is observed directly underneath the particle. 

We reproduced the gradient broadening effect with a numerical model. For a SPION located at the origin, the ensemble fluorescence signal as a function of microwave frequency, $f$, can be expressed as:
 \begin{equation}
\label{aeq:PSF_contrast}
  \begin{split}
&I(x',y',f)=\\
&I_0 {\iint\limits _{-\infty}^{~\infty}}dx dy\, [1{-}\mathscr{L}(f{-}f_{\pm},C,\Gamma)]\,\,\text{psf}(x,y)\,\,\delta(x'{-}x,y'{-}y),
  \end{split}
 \end{equation}
where $\mathscr{L}(f-f_{\pm},C,\Gamma)$ is a Lorentzian function centered at $f_{\pm}$ with contrast $C$ and FWHM $\Gamma$, $\delta(\vec{r})$ is the the delta function, and $\text{psf}(x,y)$ is the optical point-spread function. Note that the positions of the $f_{\pm}$ resonances depend on the $(x',y')$ position due to the spatial variation of the SPION field, $B_{\rm nv}$, see Eq.~\eqref{eqn:magnetic field Bnv}. As in the main text, we have assumed that the NV centers lie in a plane $z=40~{\rm nm}$ from the SPION center. 
 
We used Eq.~\eqref{aeq:PSF_contrast} to simulate the impact of SPION magnetic-field gradients on ODMR spectra. Figure~\ref{fig:sup10} shows simulated $xy$ maps of the ODMR contrast and FWHM for the confocal ($\tau_D=0$) and CSD ($\tau_D=4~{\rm \upmu s}$ cases side-by-side with the experimental maps. The agreement with experiment is fairly good, lending confidence that the features are indeed due to gradient broadening from the SPION magnetic field. Future experiments may avoid this effect by studying samples with lower magnetization and/or smaller particle sizes.

\section{Ferromagnetic nanoparticles}
\label{sec:SI micromod}
\begin{figure}[htb]
\centering
\includegraphics[width=0.95\textwidth]{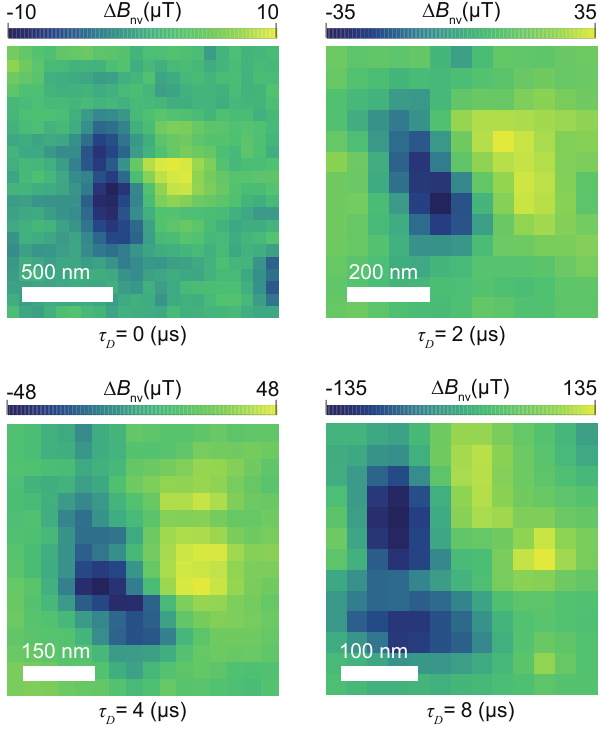}\hfill
\caption{\label{fig:micromod}
\textbf{CSD magnetic microscopy of a ferromagnetic nanoparticle.} Magnetic images of a $\sim30~{\rm nm}$ ferromagnetic nanoparticle for four different Donut pulse lengths $\tau_D=0,2,4,{\rm and~ 8~\upmu s}$. The images were acquired without the blue-reflectance feedback (\ref{sec:SI Tracking}), so they are slightly more blurry and distorted than those in the main text.}
\end{figure}

We also used CSD magnetic microscopy to image the magnetic fields produced by ferromagnetic nanoparticles purchased from Micromod (Synomag and Synomag-D). Similar to the SPION sample, these nanoparticles showed a characteristic size of $\sim30~{\rm nm}$ under SEM images. We drop-cast a diluted sample on a diamond sensor, examined them with SEM, and then performed super-resolution magnetic microscopy. Figure \ref{fig:micromod} shows magnetic images for $\tau_D=0,2,4,{\rm and~ 8~\upmu s}$. The feature widths and amplitudes are comparable to those observed with the SPION samples. Unlike $\rm SPIONs$, ferromagnetic nanoparticles are expected to have a randomly-oriented magnetic moment that is not necessarily aligned to the applied field. However, for this nanoparticle, the magnetic moment happens to be nearly parallel with $\vec{B_0}$. This could be a coincidence, or it could be that this particle had been re-magnetized by the applied field. Magnetic images of a few other particles in this suspension did not show the same degree of alignment of magnetic moment with $\vec{B_0}$.

\section{Missing SPION}
\label{sec:Missed SPION}

\begin{figure}[!ht]
\centering
\includegraphics[width=0.35\columnwidth]{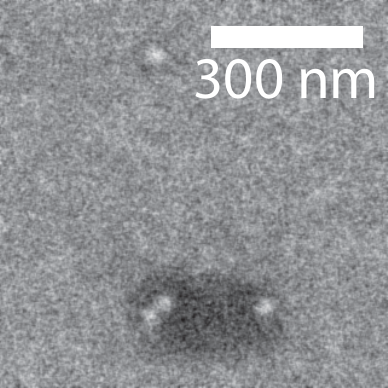}\hfill
\caption{\label{fig: SI SEM}
\textbf{Vanished SPION.} The particle in the top of the SEM image did not produce an observable magnetic signal. We suspect it was carried away from a crash with the tape used to hold the microwave wire.}
\end{figure}

While we were performing the multi-SPION experiment shown in Figure \ref{fig:multi spions}(e-f), there were inconsistencies in the signal from a SPION at the top of the field of view. The initial SEM shows a fourth SPION as in Fig.~\ref{fig: SI SEM}. During our first confocal magnetic images, we saw hints of a fourth magnetic feature in this region. However, in subsequent confocal and CSD magnetic microscopy images, we did not observe any magnetic signals from this fourth SPION. After magnetic images were taken, we inspected the region for a second time under SEM and observed that the region was contaminated with debris that obscured the view of SPIONs. We suspect that the `missing'' fourth SPION was picked up by the tape we used to keep the MW wire in place, as the loss of magnetic signal coincided with an accidental crash of the wire into the diamond during alignment. This is the only field of view we studied where such a loss of magnetic signal occurred. In all other regions, the SPION magnetic images correlated with the SEM images and did not change over time.

\section{Sensitivity model}
\label{sec:SI Sensitivity}

In order to describe the ODMR parameters and sensitivity reported in Fig.~\ref{fig:sensitivity}, we developed a model that incorporated a constant background and the ``light-narrowing effect''~\cite{DRE2011,JEN2013}. The light-narrowing effect arises from a competition between optical spin-polarization and microwave depolarization rates. At relatively high microwave power, increasing in the optical pumping rate leads to a narrowing of the ODMR line and, eventually, a reduction in contrast.

\begin{figure}[!ht]
\centering
\includegraphics[width=0.65\columnwidth]{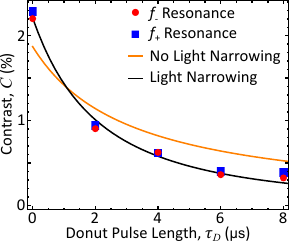}\hfill
\caption{\label{fig:contrast light narrowing}\textbf{Contrast model comparison.} ODMR contrast versus $\tau_{D}$. Red circles are the experimental contrast for the $f_-$ transition and blue squares are for $f_+$. The black curve is a fit to the model including background fluorescence and light narrowing, Eq.~\eqref{aeqn:contrast}. The orange curve is a fit with the light-narrowing term omitted, Eq.~\eqref{aeqn:contrastNOLN} }
\end{figure}

For the fluorescence rate, we need only consider the constant background and fit the data using Eq.~\eqref{eqn:PLdecay}, as discussed in the main text. For the ODMR contrast, we tried two different models. The first model assumes only a constant background that is independent of $\tau_D$ and the microwave frequency. The fit function is:
\begin{equation}
\label{aeqn:contrastNOLN}
C = \frac{C_{0}}{1 + \frac{bg}{I_0(0)} (1 + \tau_{D} / \tau_{\rm sat})},
\end{equation}
where $C_0$ is a fit parameter and $I_0(0) = 642~{\rm kcps}$ and  $bg = 162~{\rm kcps}$ are fixed at the values obtained from the $I_0(\tau_D)$ fit in Eq.~\eqref{eqn:PLdecay}. Figure~\ref{fig:contrast light narrowing} shows the experimental contrast for both $f_{\pm}$ transitions and the fitted curve using Eq.~\eqref{aeqn:contrastNOLN} in orange. The lack of agreement suggests that there is an additional mechanism causing the contrast to decrease with increasing $\tau_D$.

For the second model, we incorporate both the constant background and another term due to the light-narrowing effect (previously studied for green excitation and no Donut beam~\cite{DRE2011,JEN2013}). Here, the fit function is:
\begin{equation}
\label{aeqn:contrast}
C = \frac{C_{0}}{1 + \frac{bg}{I_0(0)} (1 + \tau_{D} / \tau_{\rm sat})} \frac{1}{1 + x~\tau_{D}},
\end{equation}
where $C_{0} = 2.8 \pm 0.1\%$ is the fitted contrast at $\tau_{D}=0$. The last term is due to light narrowing, where $x = 0.17 \pm 0.02~{\rm \upmu s}^{-1}$ is a fitted value related to the NV transverse spin relaxation rate, longitudinal spin relaxation rate, optical pumping rate, and microwave Rabi frequency. As with the first model, we held $I_0(0) = 642~\rm{kcps}$ and $bg = 162~\rm{kcps}$ at the values found from the fit in Eq.~\eqref{eqn:PLdecay}. Figure~\ref{fig:contrast light narrowing} shows the result of this fit in black, showing good agreement with our experiments. The ideal curve, shown as dashed green in Fig.~\ref{fig:sensitivity}(b), sets $bg=x=0$.

The presence of the light-narrowing effect with increasing $\tau_D$ was somewhat unexpected, since an ideal Donut beam shouldn't excite the NV$^-$ centers in the core that contribute to the Readout ODMR signal. However, the clear decrease in ODMR FWHM for increasing $\tau_D$, seen in Fig.~\ref{fig:sensitivity}(c), provides further evidence for the presence of light narrowing. We suspect that imperfect Donut intensity contrast leads to a small optical pumping rate within the Donut core, which drives the light-narrowing effect. The fit function used in Fig.~\ref{fig:sensitivity}(c) is an empirical version of that derived in Ref.~\cite{JEN2013}:
\begin{equation}
\label{aeqn:linewidth}
\Gamma = x_0 + x_1 \sqrt{\frac{1}{1 + x_2~\tau_{D}}}.
\end{equation}
Here $x_0 = 6.8\pm1.3~{\rm MHz}$ is a fitted value that depends on the inhomogeneous broadening, $x_1 = 4.2\pm1.3~{\rm MHz}$ is a fitted value that depends on the NV transverse and longitudinal spin relaxation rates and microwave Rabi frequency, and $x_2 = 0.87\pm0.75~{\rm \upmu s}^{-1}$ is a fitted value that depends on the optical pumping rate. The ideal curve, shown as dashed green in Fig.~\ref{fig:sensitivity}(c), sets $x_2=0$.

\section{Wavelength of Readout beam}
\label{sec:SI multi wavelength}

Previous CSD microscopy works used $\sim590~{\rm nm}$ orange wavelength beams for NV$^-$ Readout~\cite{CHE2015,LI2020,CHE2019,DU2020,GAR2022}. This wavelength is convenient in part because it reduces spectral overlap between second-order Raman emission (typically $80\mbox{-}110~{\rm nm}$ redshift from excitation) and the NV$^-$ collection band ($650\mbox{-}750{\rm nm}$). It also lies above the NV$^0$ zero-phonon line ($575~{\rm nm}$), suggesting that it is inefficient at converting NV$^0$ to NV$^-$. However, our experiments suggest that $637~{\rm nm}$ is a superior choice of Readout wavelength for CSD magnetic microscopy with high-density, shallow-implanted NV ensembles.

To test the CSD microscope's dependence on Readout wavelength, we first recorded ODMR spectra for different values of $\tau_D$, using 590, 600, 610, and $637~{\rm nm}$ Readoiut wavelengths. Working in a region free of magnetic features, and using the same CSD microscope as in Figure \ref{fig:experimental setup}(a) we adjusted the Readout wavelength using a supercontinuum laser~\cite{SIL2019} with an Acousto-Optic Tunable Filter (TEAF3-0.45-0.7-S-MSD, Brimrose). 

\begin{figure}[ht!]
\centering
\includegraphics[width=0.6\columnwidth]{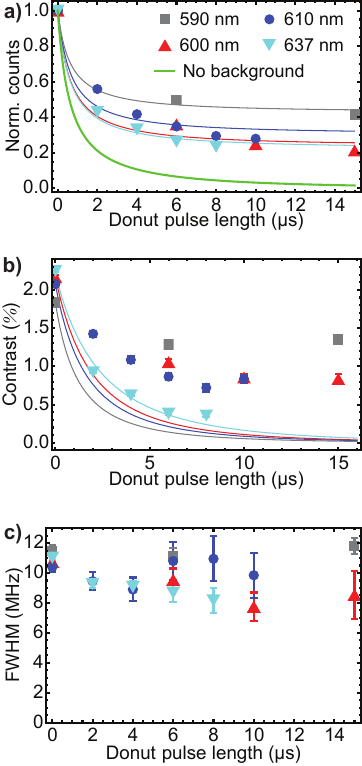}\hfill
\caption{\label{fig:sens different readouts}
\textbf{ODMR parameters for different Readout wavelengths.} (a) Fluorescence countrate, (b) Contrast, and (c) FWHM versus $\tau_D$ for different Readout wavelengths. 
}
\end{figure}

\begingroup
\setlength{\tabcolsep}{3pt}
\renewcommand{\arraystretch}{1.0}
\begin{table}[htb]
\centering
\begin{tabular}{c| c c} 
 Wavelength & Countrate at $\tau_D{=}0$, $I_0(0)$ & Background, $bg$ \\ [0.2ex] 
\hline
 590 nm & 979 kcps & 724 kcps \\ 
 600 nm & 768 kcps & 236 kcps \\
 610 nm & 895 kcps & 365 kcps \\
 637 nm & 635 kcps & 169 kcps \\ [0ex]
 \hline
\end{tabular}
\caption{\label{table:dif colors pl fit}
\textbf{Fit parameters for $I_0(\tau_D)$.} Fit parameters extracted from fitting data in Fig.~\ref{fig:mag different readouts}(a) to Eq.~\eqref{eqn:PLdecay}.}
\end{table}
\endgroup

\begin{figure*}[ht]
\centering
\includegraphics[width=0.68\textwidth]{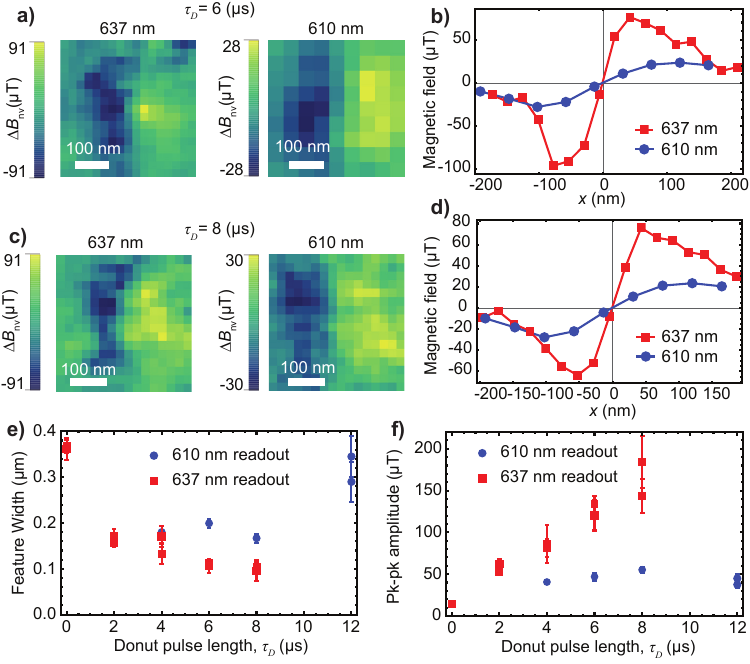}\hfill
\caption{\label{fig:mag different readouts}
\textbf{Comparison of CSD magnetic microscopy with $610~{\rm nm}$ and $637~{\rm nm}$ Readout.} (a) CSD magnetic images of a single SPION at $\tau_D=6~{\rm \upmu s}$ for two different Readout wavelengths, $637~{\rm nm}$ and $610~{\rm nm}$, and horizontal linecuts (b). (c) CSD magnetic images of the same SPION at $\tau_D=8~{\rm \upmu s}$ for $637~{\rm nm}$ and $610~{\rm nm}$ Readout, and horizontal linecuts (d). (e) Magnetic feature width versus $\tau_D$ for $637~{\rm nm}$ and $610~{\rm nm}$ Readout. (f) Magnetic feature amplitude versus $\tau_D$ for the two wavelengths.}
\end{figure*}

Figure~\ref{fig:sens different readouts}(a) shows the fluorescence countrate versus $\tau_D$ for four Readout wavelengths. Each trace is fit to Eq.~\eqref{eqn:PLdecay} to obtain $I_0(0)$ and $bg$, shown in Tab.~\ref{table:dif colors pl fit}. The fractional background fluorescence, $bg/I_0(0)$, increases as the wavelength decreases from $637$ to $590~{\rm nm}$. 

If this background were due to emission sources other than NV$^-$, then the contrast for $590\mbox{-}610~{\rm nm}$ Readout wavelengths should decrease more rapidly with increasing $\tau_D$ than the decrease observed with $637~{\rm nm}$ Readout. However, we observe the opposite behavior, Figure~\ref{fig:sens different readouts}(b). The contrast for $590\mbox{-}610~{\rm nm}$ Readout wavelengths does decrease somewhat with increasing $\tau_D$, but not nearly as rapidly as the decrease observed for $637~{\rm nm}$ Readout. We fit each Readout wavelength's $C(\tau_D)$ curve with Eq.~\eqref{aeqn:contrast}, holding $x_1$ and $x_2$ at the values observed for $637~{\rm nm}$ and fixing $bg$ and $I_0(0)$ at the values obtained from the fits in Figure~\ref{fig:sens different readouts}(a). We find very poor agreement with experiment for wavelengths below $637~{\rm nm}$. The relative lack of decay in contrast suggests that the ``background'' fluorescence is in fact mostly due to NV$^-$ centers. We suspect that the shorter Readout wavelengths are converting NV$^0$ in the Donut shell back to NV$^-$, despite the wavelength being somewhat longer than the NV$^0$ zero-phonon line wavelength. Figure~\ref{fig:sens different readouts}(c) shows the ODMR FWHM as a function of $\tau_D$. The behavior is similar for all Readout wavelengths, providing further evidence that the contrast curves cannot be explained by dramatic changes in the light-narrowing effect.

The data in Fig.~\ref{fig:sens different readouts} for $610~{\rm nm}$ Readout wavelength showed the largest drop in contrast (though still not nearly as large as expected), so we decided to try $610~{\rm nm}$ Readout in CSD magnetic microscopy and compare the results with the $637~{\rm nm}$ Readout used throughout the main text. Figure~\ref{fig:mag different readouts}(a) shows CSD magnetic images for $637~{\rm nm}$ and $610~{\rm nm}$ Readout at $\tau_D=6~{\rm \upmu s}$. The corresponding horizontal linecuts are shown in Fig.~\ref{fig:mag different readouts}(b). The $610~{\rm nm}$ Readout image does show some signs of resolution narrowing and magnetic feature amplitude enhancement, but the effects are not nearly as large as for $637~{\rm nm}$ Readout. Figure~\ref{fig:mag different readouts}(c) shows CSD magnetic images for $\tau_D=8~{\rm \upmu s}$ and Fig.~\ref{fig:mag different readouts}(d) shows horizontal linecuts for the same Readout wavelengths. The trend is the same.

Figure~\ref{fig:mag different readouts}(e) shows the feature width for both Readout wavelengths as a function of $\tau_D$. While the feature width with $637~{\rm nm}$ Readout reaches $101\pm13~{\rm nm}$ for $\tau_D=8~{\rm \upmu s}$, the width with $610~{\rm nm}$ Readout only reaches ${\sim}160~{\rm nm}$ at $\tau_D=8~{\rm \upmu s}$, and then it increases to $\gtrsim300~{\rm nm}$ for $\tau_D=12~{\rm \upmu s}$. Figure~\ref{fig:mag different readouts}(f) shows the magnetic feature amplitude for both Readout wavelengths as a function of $\tau_D$. Once again, the $637~{\rm nm}$ Readout outperforms the $610~{\rm nm}$ readout, reaching $156\pm17~{\rm \upmu T}$ at $\tau_D=8~{\rm \upmu s}$, a factor of $\sim3$ larger than for $610~{\rm nm}$ Readout. The saturation behavior of the $610~{\rm nm}$ Readout curves in Figs.~\ref{fig:mag different readouts}(e,f) is consistent with the idea that the Readout beam is not effective at keeping NV centers in the Donut shell in the NV$^0$ charge state. An increase in $\rho_{\rm shell}/\rho_{\rm core}$ leads to a smaller weight for the super-resolved image and a larger weight for the background confocal image, see~\ref{sec:SI Magnetic simulation}. This would explain why the $610~{\rm nm}$ and lower Readout wavelengths have less resolution narrowing and smaller feature amplitude enhancement.

\pagebreak
%

\end{document}